\documentclass{article} % For LaTeX2e
\usepackage{iclr2026_conference,times}

\usepackage{amsmath,amsfonts,bm}

\def\eqref#1{equation~\ref{#1}}
\def\1{\bm{1}}

\DeclareMathAlphabet{\mathsfit}{\encodingdefault}{\sfdefault}{m}{sl}
\SetMathAlphabet{\mathsfit}{bold}{\encodingdefault}{\sfdefault}{bx}{n}

\usepackage{hyperref}
\usepackage{url}
\usepackage{enumitem}
\usepackage{graphicx}
\usepackage{caption}
\usepackage{subcaption}
\usepackage{booktabs}
\usepackage{xcolor}
\usepackage{colortbl}
\usepackage{tabularx}
\usepackage{array}
\usepackage{float}
\usepackage{lipsum}
\usepackage{comment}
\usepackage{multirow}
\usepackage{diagbox}
\usepackage{threeparttable}
\usepackage{pifont} 
\usepackage[table,xcdraw]{xcolor}
\usepackage{titlesec} % 用于调整标题格式

\newcommand{\cmark}{\ding{51}} % Checkmark
\newcommand{\xmark}{\ding{55}} % Cross

\title{\includegraphics[width=0.8cm]{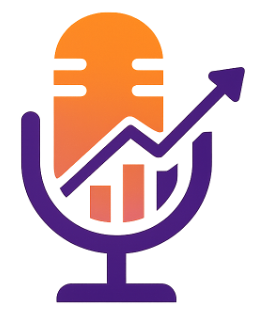} PodEval: A Multimodal Evaluation \\ Framework for Podcast Audio Generation}

\author{\textbf{Yujia Xiao}$^{1}$, \textbf{Liumeng Xue}$^{2}$, \textbf{Lei He}$^{3}$, \textbf{Xinyi Chen}$^{4}$, \textbf{Aemon Yat Fei Chiu}$^{1}$, \textbf{Wenjie Tian}$^{5}$, \\
\textbf{Shaofei Zhang}$^{3}$\textbf{,} \textbf{Qiuqiang Kong}$^{1}$\textbf{,} \textbf{Xinfa Zhu}$^{5}$\textbf{,} \textbf{Wei Xue}$^{2}$\textbf{,} \textbf{Tan Lee}$^{1,\dagger}$\\
$^{1}$The Chinese University of Hong Kong, Hong Kong, China\\
$^{2}$The Hong Kong University of Science and Technology, Hong Kong, China\\
$^{3}$Microsoft, China\\
$^{4}$South China University of Technology, China\\
$^{5}$Northwestern Polytechnical University, China\\
\texttt{yujiaxiao@link.cuhk.edu.hk}\\
\texttt{tanlee@ee.cuhk.edu.hk}\\
}

\begin{document}

\maketitle

{
\makeatletter
\def\@fnsymbol#1{\ensuremath{\ifcase#1\or *\or \dagger\or **\or ***\or \ddagger\else\@ctrerr\fi}}
\makeatother
\renewcommand{\thefootnote}{\fnsymbol{footnote}}
%\footnotetext[1]{First Authors.}
\footnotetext[2]{Corresponding Author.}
}
\setcounter{footnote}{0}
\renewcommand{\thefootnote}{\arabic{footnote}}

\begin{abstract}

\vspace{-0.2cm}

Recently, an increasing number of multimodal (text and audio) benchmarks have emerged, primarily focusing on evaluating models' understanding capability. However, exploration into assessing generative capabilities remains limited, especially for open-ended long-form content generation. Significant challenges lie in no reference standard answer, no unified evaluation metrics and uncontrollable human judgments. In this work, we take podcast-like audio generation as a starting point and propose PodEval, a comprehensive and well-designed open-source evaluation framework. In this framework: 1) We construct a real-world podcast dataset spanning diverse topics, serving as a reference for human-level creative quality. 2) We introduce a multimodal evaluation strategy and decompose the complex task into three dimensions: text, speech and audio, with different evaluation emphasis on ``Content" and ``Format". 3) For each modality, we design corresponding evaluation methods, involving both objective metrics and subjective listening test. We leverage representative podcast generation systems (including open-source, close-source, and human-made) in our experiments. The results offer in-depth analysis and insights into podcast generation, demonstrating the effectiveness of PodEval in evaluating open-ended long-form audio. This project is open-source to facilitate public use: \url{https://github.com/yujxx/PodEval}.
%All datasets and evaluation tools are open-sourced with clear instructions to facilitate public use: xxx.

\end{abstract}

\vspace{-0.4cm}

\section{Introduction}

\vspace{-0.2cm}

%With the rapid development of AIGC (AI-Generated Content) in recent years, many innovative applications have emerged, with AI Podcast becoming a key application scenario for audio-based generative models. Google Labs introduced NotebookLM \cite{notebooklm2023} in late 2023, initially designed to extract key insights from provided materials. However, it gained widespread popularity in late 2024 with the launch of its ``Audio Overviews" feature. This feature transforms given materials into conversational content, presented as a two-person podcast in audio format. The success of this feature demonstrated the demand for podcast-style media, leading to the emergence of similar products, such as ByteDance's ``Doubao" \cite{bytedance2025} AI Podcast Model, released in 2025.

With the rapid development of AIGC (AI-Generated Content) in recent years, many innovative applications have emerged. AI Podcast represents a key application scenario for audio-based generative models \citep{notebooklm2023,bytedance2025}. However, evaluating podcast-like audio is challenging due to: 1) it is an open-ended task, which means there is no reference standard answer; 2) the evaluation of long-form speech/audio is particularly difficult, as longer formats introduce more variability. Objective metrics often fail to capture human perceptions accurately, while subjective listening tests face issues like user inattention, which reduces the validity of results; and 3) podcasts often incorporate additional elements, like music and sound effects, making the evaluation more complicated. 

To address these challenges and establish a clear evaluation framework, we decompose podcast-like audio into three dimensions: \textbf{text} (conversation transcripts), \textbf{speech} (spoken dialogue), and \textbf{audio} (speech, music, sound effects, and their interaction). While these dimensions inherently overlap, they offers a structured framework for evaluation focus. Specifically, the conversation transcripts in podcasts are primarily used for \textbf{content} (the message being conveyed) evaluation, whereas speech, music and sound effects primarily contribute to \textbf{format} (how the message is presented) evaluation.

Different modalities have their own commonly used evaluation methods. For text, metrics such as BLEU \citep{papineni2002bleu}, ROUGE \citep{lin2004rouge}, and METEOR \citep{banerjee2005meteor} focus on fluency and relevance, while newer approaches like BERTScore \citep{zhang2019bertscore} utilize pre-trained language models to capture semantic alignment. For speech, objective metrics like Mel Cepstral Distortion (MCD) and Perceptual Evaluation of Speech Quality (PESQ) \citep{rix2001perceptual} are widely used, alongside subjective evaluations like Mean Opinion Score (MOS) \citep{international1996methods}. For audio, metrics like Frechet Audio Distance (FAD) \citep{kilgour2018fr} and Kullback-Leibler Divergence (KL) are employed to evaluate audio quality, while listener surveys provide subjective insights. However, these evaluation methods are not directly applicable to podcast evaluation since:

%\vspace{-0.2cm}

\begin{itemize}[leftmargin=10pt]
\item Most content-related objective metrics rely on reference scripts to measure quality and relevance. However, podcast generation lacks standardized references as it is an open-ended generation task. Moreover, relying on such references limits the diversity and creativity of the generated content.
%\vspace{-0.1cm}
\item General speech evaluation focuses on individual sentences, while podcasts require natural and interactive dialogue, emphasizing dialogue-level naturalness. Additionally, voice presentation in multi-speaker scenarios is critical to ensuring role distinction and overall listener engagement.
%\vspace{-0.1cm}
\item While music and sound effects are not essential to every podcast, their evaluation, when present, should go beyond the quality of individual audio events. Instead, it should focus on their overall harmony and seamless integration with the speech content to enhance the listener's experience.
%\vspace{-0.1cm}
\item Subjective tests are essential for open-ended generative tasks, but crowdsourced data often faces reliability issues, as it is difficult to control or determine whether users are attentive. Especially for long-form content, users may lose focus or respond randomly, which affects the result validity.
%as it is difficult to control or determine whether users are responding attentively. This is particularly challenging for long-form content evaluations, where users are prone to losing focus or answering randomly, thereby compromising the validity of the results.
\end{itemize}
%\vspace{-0.2cm}
%As there are numerous factors involved in constructing a podcast, the evaluation is inherently complex. 
In this work, we introduce \textbf{PodEval}, a comprehensive multimodal evaluation framework designed for podcast-like long-form audio generation. The contributions can be summarized as:

%\vspace{-0.2cm}

\begin{itemize}[leftmargin=10pt]
\item We construct a real-world podcast dataset spanning a wide range of podcast categories and topics, serving as a reference for human-level creative quality. Model-based samples are also provided.

%\vspace{-0.1cm}

\item We decompose podcast-like audio evaluation from a multimodal viewpoint—text, speech, and audio—to establish a clear evaluation framework, with distinct focuses on “Content” and “Format”. 

%\vspace{-0.1cm}

\item For each modality, we design tailored metrics to address diversity considerations. For text, we combine quantitative metrics with LLM-based evaluations to assess conversation scripts. For speech and audio, we design objective metrics and subjective listening tests to evaluate spoken dialogue and overall audio performance. All evaluation methods are organized into open-source tools for ease of use. Subjective tests are enhanced by spammer detection to improve data validity.
%Subjective listening tests are meticulously designed to enhance data validity.

%\vspace{-0.1cm}

\item We utilize representative podcast generation systems in our experiments, including open-source, closed-source, and human-made ones. The results offer detailed analyses of these systems, provide insights for podcast generation, and validate the effectiveness of our evaluation framework.
\end{itemize}

%ICLR requires electronic submissions, processed by
%\url{https://openreview.net/}. See ICLR's website for more instructions.

%If your paper is ultimately accepted, the statement {\tt
%  {\textbackslash}iclrfinalcopy} should be inserted to adjust the
%format to the camera ready requirements.

%The format for the submissions is a variant of the NeurIPS format.
%Please read carefully the instructions below, and follow them
%faithfully.
%\vspace{-0.4cm}

\section{Related work}

%\vspace{-0.2cm}

\subsection{Podcast generation}

%\vspace{-0.2cm}
%Recent advancements in generative AI for text, speech, and audio have shown remarkable potential in creating realistic and high-quality content across multiple modalities. Large language models (LLMs)\cite{ouyang2022,achiam2023gpt,team2023gemini,touvron2023llamaopenefficientfoundation, anthropic_claude} can produce coherent and contextually relevant scripts, while zero-short text-to-speech (TTS) systems \cite{casanova2022yourtts,wang2023neural,sunoai_bark,tan2024naturalspeech,lajszczak2024base,du2024cosyvoice1,du2024cosyvoice,wang2025sparkttsefficientllmbasedtexttospeech} are capable of delivering expressive and natural-sounding voices for any speaker using a brief reference sample. Additionally, text-to-audio (TTA) models \cite{kreuk2022audiogen, liu2023audioldm, liu2024audioldm, huang2023make,ghosal2023text} can synthesize background music and sound effects tailored to specific scenarios. 

Podcasts are a popular audio format, with platforms like Apple Podcasts and Spotify leading the way. The rise of the AI podcast began with Google's NotebookLM \citep{notebooklm2023}, which gained popularity in late 2024 for its ``Audio Overviews" feature. This feature converts materials into conversational, two-person podcasts, praised for its highly natural dialogue speech. Similarly, most open-source podcast generation systems focus on dialogue speech synthesis, like Dia \citep{nari-labs-dia}, Muyan-TTS \citep{li2025muyanttstrainabletexttospeechmodel}, MoonCast \citep{ju2025mooncasthighqualityzeroshotpodcast} and MOSS-TTSD \citep{moss2025ttsd}. These systems function primarily as dialogue Text-to-Speech (TTS) engines for text-given scenarios. Another type of podcast generation system takes a more holistic approach, incorporating elements beyond speech, such as text and music/sound. For example, WavJourney \citep{liu2023wavjourney} leverages LLMs to connect components like TTS and Text-to-Audio (TTA), generating element-rich audio programs. Upon this, PodAgent \citep{xiao2025podagent} introduces a ``Host-Guest-Writer" multi-agent system to create informative conversation scripts and builds a voice pool for appropriate voice selection. Table \ref{tab:podcast_systems} compares the systems leveraged in subsequent experiments.

%\vspace{-0.2cm}

\begin{table}[h]
\caption{\centering Comparison of podcast generation systems.}
\label{tab:podcast_systems}
\vspace{-0.1cm}
\centering
\setlength{\tabcolsep}{4pt} % Adjust column spacing
\resizebox{\textwidth}{!}{
\begin{threeparttable} % Wrap the table in the threeparttable environment
\begin{tabular}{lccccc}
\toprule
\rowcolor[HTML]{F2F2F2} 
\textbf{System} & \textbf{Open-Source?} & \textbf{\# Speaker} & \textbf{Support Voice Selection?} & \textbf{Is Dialogue TTS?} & \textbf{Support Music/Sound?} \\
\midrule
NotebookLM       & \xmark               & 2                   & \xmark                            & -                        & \xmark                        \\
Dia              & \cmark               & 2                   & Preset                            & \cmark                   & \xmark                        \\
Muyan-TTS        & \cmark               & 1                   & Preset                            & \xmark                   & \xmark                        \\
MoonCast         & \cmark               & 2                   & Preset                            & \cmark                   & \xmark                        \\
MOSS-TTSD        & \cmark               & 2                   & Preset                            & \cmark                   & \xmark                        \\
PodAgent\tnote{*} & \cmark & N                   & Auto                              & \xmark                   & \cmark                        \\
\bottomrule
\end{tabular}
\begin{tablenotes}
\footnotesize
\item[*] PodAgent uses CosyVoice2\citep{du2024cosyvoice} as its backend TTS model, which is a single-sentence TTS system.
\end{tablenotes}
\end{threeparttable}}
\end{table}

%\vspace{-0.3cm}

%For example, Dia, Muyan-TTS..., MoonCast..., Dia..., MOSS-TTSD.... 
%Following this, some notable open-source podcast generation systems emerged and provide valuable insights from different perspectives. WavJourney\cite{liu2023wavjourney} is the first to leverage LLMs to connect components such as TTS and TTA, generating element-rich audio programs. Building on this, PodAgent \cite{xiao2025podagent} introduces a ``Host-Guest-Writer" multi-agent system to create complete and informative content, along with a voice pool for optimal voice-role matching. Meanwhile, Mooncast \cite{ju2025mooncasthighqualityzeroshotpodcast} focuses on generating spontaneous, long-form speech in a podcast style, and Muyan-TTS \cite{li2025muyanttstrainabletexttospeechmodel} is a zero-shot TTS model trained on a large podcast audio dataset. In this work, we leverage \textbf{PodAgent}, \textbf{Mooncast}, and \textbf{Muyan-TTS} as the podcast generation systems to be evaluated.

%% Please note that we have introduced automatic line number generation
%% into the style file for \LaTeXe. This is to help reviewers
%% refer to specific lines of the paper when they make their comments. Please do
%% NOT refer to these line numbers in your paper as they will be removed from the
%% style file for the final version of accepted papers.

%\vspace{-0.1cm}

\subsection{Evaluation on generative models}

Various evaluation works have emerged along with the development of LLMs and multimodal generative models. \textbf{Text-related Evaluation}, such as SuperGLUE, MMLU, and BIG-bench \citep{wang2019superglue,hendrycks2020measuring,srivastava2022beyond}, assesses the capabilities of LLMs across diverse tasks with preset ground truth. Subsequently, MT-Bench \citep{zheng2023judging} explores the potential of LLMs as evaluators, and Chatbot Arena \citep{chiang2024chatbot} provides an open platform for assessing LLMs based on human preferences. \textbf{Speech-related Evaluation}, such as SUPERB \citep{yang2021superb}, is designed for discriminative tasks like speech recognition and speaker identification. However, evaluations for generative tasks are scarce due to their inherent diversity and subjectivity, making subjective evaluation essential for speech generation tasks. For instance, VOCBENCH \citep{albadawy2022vocbench} incorporates both subjective and objective evaluations to assess vocoder performance in speech synthesis. Similarly, numerous \textbf{Audio-related Evaluation} work, such as AIR-Bench, Audiobench, MMAU, and MMAR \citep{yang2024air,wang2024audiobench,sakshi2024mmau,ma2025mmar}, focus on audio understanding and reasoning. Subjective evaluation remains crucial for assessing audio generation systems and is typically tailored to specific generation tasks. Unlike existing evaluation works, \textbf{PodEval} introduces a comprehensive framework specifically designed for podcast-like audio generation. It emphasizes both subjective and objective evaluations across text, speech, and audio, with all metrics closely aligned with real-world user experience.
%The style files for ICLR and other conference information are available online at:
%\begin{center}
%   \url{http://www.iclr.cc/}
%\end{center}
%The file \verb+iclr2026_conference.pdf+ contains these
%instructions and illustrates the
%various formatting requirements your ICLR paper must satisfy.
%Submissions must be made using \LaTeX{} and the style files
%\verb+iclr2026_conference.sty+ and \verb+iclr2026_conference.bst+ (to be used with \LaTeX{}2e). The file
%\verb+iclr2026_conference.tex+ may be used as a ``shell'' for writing your paper. All you
%have to do is replace the author, title, abstract, and text of the paper with your own.

%The formatting instructions contained in these style files are summarized in sections \ref{gen_inst}, \ref{headings}, and \ref{others} below.

%5\vspace{-0.1cm}

\section{Real-Pod: Real-world podcast dataset}
\label{dataset}

%\vspace{-0.2cm}
\begin{figure}[h]
  \centering      
  \includegraphics[width=1\columnwidth]{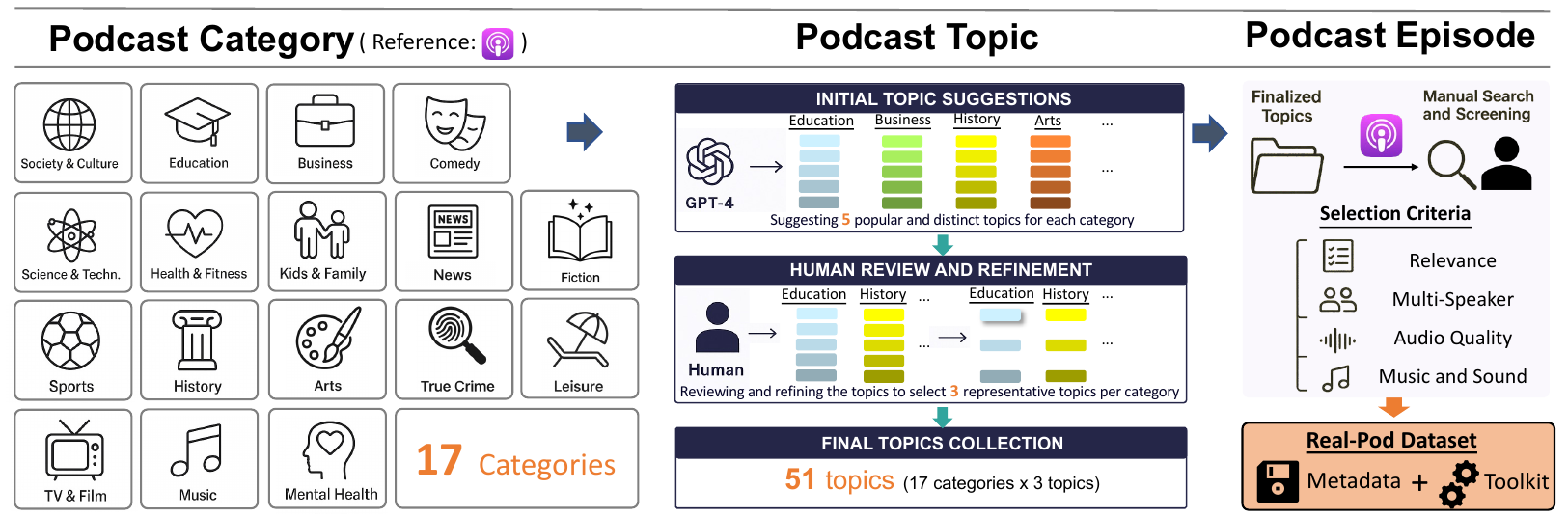}
  \captionsetup{justification=centering}
  \caption{The workflow for constructing the Real-Pod dataset.}
  \label{fig:dataset}
\end{figure}
%\vspace{-0.2cm}

There is no unified standard for defining ``what makes a good podcast episode." Unlike textbooks or official TV programs, podcasts can be created by anyone to share their unique ideas or insights. We do not make direct comparisons between generated podcasts and real podcasts—such comparisons are inherently unfeasible, especially when they approach topics from entirely different perspectives. Instead, we construct a real-world podcast dataset, called \textbf{Real-Pod} dataset, to serve as a reference for human-level creative quality. It is important to note that this dataset acts as a ``reference" rather than an absolute ``answer". The design principles of the Real-Pod dataset are \textbf{real} (consists of human-made podcasts), \textbf{broad} (diverse topic coverage) and \textbf{rich} (varied formats, like multi-speaker, music and sound). The workflow for constructing the Real-Pod dataset is illustrated in Figure \ref{fig:dataset}:

%\vspace{-0.2cm}

\begin{itemize}[leftmargin=10pt]
\item \textbf{Podcast Category}. We began by compiling a comprehensive list of podcast categories based on the taxonomy from Apple Podcast \citep{applepodcasts}. The 17 categories are shown in Figure \ref{fig:dataset}-left.
%These 17 categories include: Society \& Culture, Education, Business, Comedy, Science \& Technology, Health \& Fitness, Arts, News, Sports, History, Kids \& Family, True Crime, TV \& Film, Music, Leisure, Fiction, Mental Health. 

\item \textbf{Podcast Topic}. Next, we established relevant topics for each category through a two-step process: (1) using GPT-4 \citep{achiam2023gpt} to generate 5 popular and distinct topics per category, reflecting current trends and listener interests; and (2) manually reviewing and refining these topics to ensure their uniqueness and relevance with real-world podcasts, selecting 3 representative topics for each category, resulting in a final collection of 51 topics (17 categories × 3 topics).

\item \textbf{Podcast Episode}. After finalizing the topic collection, we manually searched and screened podcast episodes to identify those most relevant to the selected topics. The selection process was guided by: (1) Topic Relevance: Episodes were selected based on their alignment with the predefined topics. (2) Rich Format: Preference was given to episodes that featured multi-speaker conversations, included background music and sound effects, and exhibited high audio quality.

\end{itemize}

\vspace{0.1cm}

\section{Text-based evaluation}
\label{text}
%\vspace{-0.3cm}

The dialogue content in podcasts is extracted in text format for evaluation, representing the core message the podcast aims to convey. Podcast dialogues often center around specific topics, showcasing participants' unique perspectives and insights, which makes reference-correlation-based methods infeasible. Instead, the richness of perspectives conveyed (to provide informative takeaways for the listener) and the presentation style of the dialogue (to enhance listener comprehension) should be the primary focus of evaluation. Therefore, we follow the dialogue script-based evaluation methods proposed in PodAgent \citep{xiao2025podagent}, which adopt a two-fold approach: (1) \textbf{Quantitative Metrics} such as Distinct-N, Semantic-Div, MATTR, and Info-Dens to assess lexical diversity, semantic richness, vocabulary richness, and information density, respectively. These metrics operate independently of reference texts and focus on intrinsic text characteristics; (2) \textbf{LLM-as-a-Judge}, leveraging GPT-4 to replace human evaluators for complex and comprehensive assessments. Evaluation criteria include coherence, engagingness, diversity, informativeness and speaker diversity. It incorporates comparative evaluations to reduce bias and evidence-based scoring for robust and reliable results.

\vspace{0.1cm}

\section{Speech-based evaluation}
\label{speech}
%\vspace{-0.2cm}

Speech is the core component of a podcast, serving as the medium for content delivery, and how the message delivered plays a crucial role in shaping the listening experience. To ensure a multidimensional evaluation, we first integrate the following \textbf{Objective Metrics}: 

%\vspace{-0.2cm}

\begin{itemize}[leftmargin=10pt]

\item  \textbf{WER} (Word Error Rate) measures pronunciation accuracy, a critical indicator of the robustness of TTS-based podcast generation systems, powered by Whisper \citep{radford2022robustspeechrecognitionlargescale} in our toolkit.

%\vspace{-0.1cm}

\item  \textbf{DNSMOS} \citep{reddy2022dnsmos} evaluate the speech quality (SIG), background noise quality (BAK), and overall quality (OVRL, P808\_MOS) of speech. SIG, BAK, and OVRL are trained according to P.835 \citep{recommendation2003subjective}, while P808\_MOS is trained based on P.808 \citep{itu2018p808}.

%\vspace{-0.1cm}

\item \textbf{SIM} stands for Speaker Similarity. In podcast generation systems, zero-shot TTS is often employed to replicate the voice of a preset speaker. The SIM between the synthesized voice and the reference voice serves as a crucial metric about vocal fidelity. In PodEval, SIM is quantified using the cosine similarity of extracted speaker embeddings \citep{Plaquet23, Bredin23}.

%\vspace{-0.1cm}

\item \textbf{SPTD} is a brand new metric we proposed, standing for Speaker Timbre Difference. As audio programs, podcasts are accessible only through listening. In multi-speaker conversations, voices with greater timbre differences enhance clarity and make the information easier to follow. SPTD is to assess the overall timbre variation across speakers. Equation \ref{sptd} calculates the SPTD among \( N \) distinct speakers.
%stands for Speaker Timbre Difference. As audio programs, podcasts are accessible only through listening. In multi-speaker conversations, voices with greater timbre differences enhance clarity and make the information easier to follow. Therefore, we define SPTD to assess the overall timbre variation across speakers. Equation \ref{sptd} calculates the SPTD among \( N \) distinct speakers.

\end{itemize}

%\vspace{-0.6cm}

\begin{equation}
\label{sptd}
\text{SPTD} = 1 - \frac{2}{N (N-1)} \sum_{i=1}^{N} \sum_{j=i+1}^{N} \text{sim}(\mathbf{e}_i, \mathbf{e}_j)
\end{equation}

%\vspace{-0.2cm}

Objective metrics can be calculated efficiently at a low cost without human involvement. However, the \textbf{Subjective Listening Test} remains a necessary indicator of human perception. Unlike general speech synthesis, which emphasizes sentence-level pronunciation accuracy and naturalness, podcast speech focuses on achieving human-like natural dialogue. Subjective tests for such long-form speech present several key \textbf{challenges}: \textcolor{blue}{1)} the length of dialogue in podcasts ranges from a few minutes to over an hour, making it impractical to evaluate the entire speech directly; \textcolor{blue}{2)} the difficulty of comparing more than two systems simultaneously; \textcolor{blue}{3)} guiding user focus toward dialogue naturalness, rather than on factors like content; \textcolor{blue}{4)} balancing topic diversity within a fixed testing capacity; and \textcolor{blue}{5)} ensuring that crowdsourced evaluators remain focused and provide reliable feedback.

In PodEval, we design the \textbf{Dialogue Naturalness Evaluation} based on the MUSHRA framework \citep{schoeffler2018webmushra}. The key insight from this framework, \textit{incorporating both high-quality and low-quality anchors}, helps evaluators establish a reliable reference of quality range. For researchers, analyzing scores for these anchors helps identify inattentive evaluators, enabling the \textit{filtering of invalid submissions} and improving the data vadility. In our task, we use real podcast segments from the \textit{Real-Pod dataset as the high-quality anchor} and synthesized dialogue segments from \textit{\cite{espeak} as the low-quality anchor}. For podcast samples from different systems, we provide an automatic toolkit to extract dialogue segments featuring \textit{turn-taking} between speakers, representing a typical dialogue flow. Each dialogue segment is extracted with a \textit{preset length} (e.g. 15–25 seconds) to ensure the speech samples are of similar duration. We select dialogue segments from \textit{all 17 categories} in the Real-Pod dataset to ensure content diversity while keeping the total listening test duration \textit{within 30 minutes}. In each test group, samples from different systems are presented \textit{on the same page}, along with a \textit{reference Real-Pod sample} to guide evaluators on what a natural dialogue sounds like. The scoring is adjusted using a slider ranging from 0 to 100, divided into \textit{five stages with a clear definition}. Detailed instructions and website design can be found in Appendix \ref{apdx:speech-sbj}. \footnote{The demo website is hosted at \url{https://podeval.github.io/PodEval-Subjective/?config=dialogue.yaml}. Everyone is welcome to try it out and view the results at the end.}

%Citations within the text should be based on the \texttt{natbib} packageand include the authors' last names and year (with the ``et~al.'' construct for more than two authors). When the authors or the publication are included in the sentence, the citation should not be in parenthesis using \verb|\citet{}| (as in ``See \citet{Hinton06} for more information.''). Otherwise, the citation should be in parenthesis using \verb|\citep{}| (as in ``Deep learning shows promise to make progress towards AI~\citep{Bengio+chapter2007}.'').

%The corresponding references are to be listed in alphabetical order of authors, in the \textsc{References} section. As to the format of the references themselves, any style is acceptable as long as it is used consistently.

\section{Audio-based evaluation}
\label{audio}
\vspace{-0.1cm}

In this section, we introduce the audio-based evaluation for podcasts, which treats speech as one component and assesses the overall audio performance, including speech, music and sound effects (MSE), and their interactions. Similarly, we first introduce the following \textbf{Objective metrics}:
\vspace{-0.1cm}

\begin{itemize}[leftmargin=10pt]

%\item \textbf{Loudness}: Loudness is an indicator that determines whether audio falls within an acceptable volume range. The ITU-R BS.1770-4 standard \cite{series2011algorithms} is a widely recognized framework for measuring audio loudness and true-peak levels. Building upon this, the \cite{ebu2011loudness} standard has been broadly adopted by broadcasting and streaming platforms. According to it, the target Integrated Loudness (LOUD-IT) is -23 LUFS (±1 LUFS), the True Peak (LOUD-TP) should not exceed -1 dBTP and the recommended Loudness Range (LOUD-RA) is under 20 LU. For podcast-like online streaming, adjustments are made to account for typical listening environments, such as streaming platforms or mobile devices where headphones are commonly used. In these cases, the LOUD-IT is typically recommended to be -18, -16 (±1) or -14 LUFS \cite{AES2021Loudness, ApplePodcastsAudioRequirements, SpotifyPodcastAdsRequirements}, while Netflix recommends keeping the LOUD-RA between 4 and 18 LU \cite{NetflixSoundMix2024}. There is no single ``absolute right" reference for loudness metrics. We propose the following reference standards by considering all above guidelines: \textbf{LOUD-IT:} \(-18\) to \(-14\) LUFS; \textbf{LOUD-TP:} \(\leq -1\) dBTP; \textbf{LOUD-RA:} 4 to 18 LU. Based on this ``relatively correct" reference range, we can analyze the distribution of loudness metrics across different systems. We also provide a quantitative score calculation strategy in Appendix \ref{ap_obj} for your reference.
\item \textbf{Loudness}: Loudness ensures audio falls within an acceptable volume range. The ITU-R BS.1770-4 standard \citep{series2011algorithms} is widely recognized for measuring audio loudness and true-peak levels. Based on this, the \citep{ebu2011loudness} standard has been broadly adopted by broadcast and streaming platforms,  recommend a target Integrated Loudness (LOUD-IT) of -23 LUFS (±1 LUFS), True Peak (LOUD-TP) \(\leq -1\) dBTP and Loudness Range (LOUD-RA) \(< 20\) LU. For podcast-like streaming, adjustments are made for typical listening environments, such as mobile devices where headphones are commonly used. In these cases, the LOUD-IT is recommended as -18, -16 (±1) or -14 LUFS \citep{AES2021Loudness, ApplePodcastsAudioRequirements, SpotifyPodcastAdsRequirements}. Netflix recommends keeping LOUD-RA between 4 and 18 LU \citep{NetflixSoundMix2024}. There is no ``absolute right" reference for loudness metrics. We propose the following reference standards considering all above guidelines: \textbf{LOUD-IT:} \(-18\) to \(-14\) LUFS; \textbf{LOUD-TP:} \(\leq -1\) dBTP; \textbf{LOUD-RA:} 4 to 18 LU. Based on this ``relatively correct" reference, we can analyze the distribution of loudness metrics across different systems. We also provide a quantitative scoring strategy in Appendix \ref{ap_obj}.

%The detailed score calculation strategy is provided in Appendix \ref{ap_obj}.
%ensuring natural vocal representation while minimizing the need for frequent volume adjustments by listeners

\item \textbf{SMR} (Speech-to-Music Ratio): MSE are typically integrated into podcast audio to enhance the overall listening experience. Since speech is the primary focus in podcasts, it is essential to ensure that MSE dose not overpower or mask the speech, maintaining clarity and intelligibility of the dialogue. SMR measures the balance between speech and MSE, with a minimum requirement of being greater than 0. SMR\_SCORE is the proportion of cases where SMR exceeds 0.

\item \textbf{CASP} (MSE-Speech Harmony): Harmony between speech and MSE is an advanced requirement. Appropriate MSE can enhance audio engagement, while discordant MSE distracts and negatively impacts the experience. The DualScore, calculated by the CASP framework proposed in \cite{tian2025dualdub}, measures the correlation between audio and speech. In PodEval, we employ the CASP model, pretrained on $\sim$1,000 hours of podcast data, to assess MSE-Speech Harmony.

\end{itemize}

\textbf{Subjective Listening Test} is primarily designed based on the perceptions of real users. A key challenge lies in how to evaluate extra-long audios. As we mentioned above, podcasts in the real world range from a few minutes to over an hour in length. Conducting listening tests on full-length podcast episodes is impractical due to the time, effort, and financial resources required. Moreover, it is hard to judge podcasts of vastly different lengths in a fair and consistent manner. Research on long-form audio evaluation is limited. \cite{clark2019evaluating} did investigation on long-form speech evaluation and found that multiple evaluations are necessary due to the low correlation observed across different experimental settings. \cite{cambre2020choice} conducted a comprehensive evaluation of voice selection for long-form content; however, the minimum required listening time was only 10 seconds. A podcast-related evaluation study \citep{austria2007developing} designed a questionnaire with carefully crafted questions in terms of both content and presentation to assess domain-specific podcasts. Different from that, PodEval does not constrain the domain of podcasts, and open-ended content evaluation is separately conducted in the text-based evaluation section. In this audio-based evaluation, we focus on assessing the overall performance of the audios. The design approach is as follows:

\begin{itemize}[leftmargin=10pt]

\item We design it as a \textbf{MOS test}, where evaluators listen to one audio sample at a time and provide judgments based on predefined criteria. Compared to comparative methods, this approach is more suitable for long-form content by avoiding attention overload and consistency compromising. 
\vspace{-0.1cm}
%Focusing on one sample at a time better preserves the authenticity of immediate perception.

\item The test data are preprocessed by extracting \textbf{the first / middle / final minute}. These segments are concatenated into a single audio, separated by a beep signal. This method unifies podcast length, captures overall performance from diverse positions, and minimizes content-related biases.
\vspace{-0.1cm}

\item The judgment session consists of a \textbf{questionnaire} with 8 questions covering multiple dimensions, integrating both perceptual (e.g., ``Information Delivery Effectiveness") and preference-based (e.g., ``Speaker Expression Preference") questions. This distinction helps clarify whether the ratings are rooted in objective perception or subjective preference. Users are also asked about their willingness to listen to the full episode and the perceived human likelihood, offering insights into interest levels and audio naturalness. The detailed content can be found in Appendix \ref{appendix:Audio-sbj-final}.
\vspace{-0.1cm}

\item We implement two strategies to enhance the validity of the collected data. \textbf{1) Attention-check questions:} These include questions like \textit{Q1. How many speakers are there in the podcast?} and \textit{Q7. If music or sound effects... (Select Neutral if none are present)}. These questions have standard answers, allowing us to determine whether users are actively listening to the audio. \textbf{2) Justification for answers:} Users have to provide justifications for their responses to each question, which can be short but are required. This requirement significantly increases users' focus and we can collect more detailed information from their justification. By employing these two strategies, data validity is enhanced by promoting attentiveness and filtering out unreliable responses.

\end{itemize}

\section{Experiments}
\vspace{-0.1cm}

\subsection{Text-based evaluation}
\vspace{-0.1cm}

The text-based evaluation is conducted among GPT-4, PodAgent, MoonCast, and Real-Pod. Other systems in Table \ref{tab:podcast_systems} are excluded as they do not provide conversation scripts. PodAgent, with its Host-Guest-Writer multi-agent system, can directly generate podcast scripts based on a given topic. While MoonCast functions similarly to NotebookLM, requiring external knowledge sources but providing prompt template for spontaneous script generation. For this evaluation, the MoonCast system uses the podcast scripts generated by PodAgent as input and transforms them into a spontaneous version.

\begin{figure}[h]
  \centering      
  \includegraphics[width=1\columnwidth]{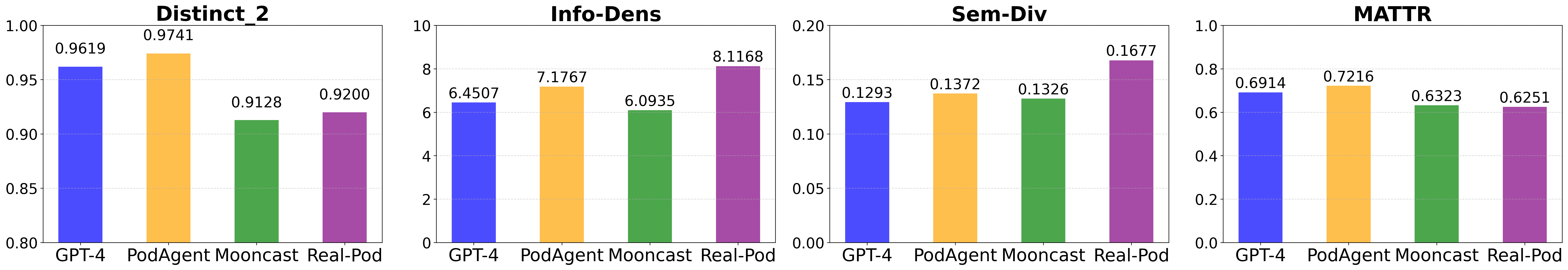}
  \captionsetup{justification=centering}
  \caption{\textbf{Quantitative metrics:} comparison among GPT-4, PodAgent and Real-Pod.}
  \label{fig:text-1}
\end{figure}

\textbf{Quantitative Metrics.}   Detailed scores calculated across 17 podcast categories for each system are presented in Appendix \ref{appendix:quant_text}. For a concise and clearer comparison, we present the overall performance (averaged across all 17 categories) in Figure \ref{fig:text-1}, where we can observe that: \textcolor{blue}{1)} For each quantitative metric, PodAgent outperforms directly prompting GPT-4; \textcolor{blue}{2)} When comparing LLM-based methods (GPT-4, PodAgent) with human-created podcasts (Real-Pod), Real-Pod scores lower on lexical diversity (Distinct-2 and MATTR) but higher on information density and semantic diversity (Info-Dens and Sem-Div). This is reasonable for: i) real human interactions often include filler words and use simpler language; ii) most real podcasts are significantly longer (30 minutes to an hour), leading to higher information richness compared to generated podcasts, which are usually only a few minutes long; \textcolor{blue}{3)} As a spontaneous version of PodAgent, MoonCast shows reduced lexical diversity and information density. While its semantic diversity remains comparable to PodAgent.

%\vspace{-0.1cm}

\textbf{LLM-as-a-Judge.}   This evaluation compares PodAgent (scored from -3 to 3) with GPT-4 (reference score as 0), both of which generate conversation scripts without external knowledge resources. Detailed scores for each category are provided in Appendix \ref{appendix:LLM_text}. We present the overall performance and results for five specific categories in Table \ref{tab:llm-judge-part} for analysis. We can see that scores across all metrics and all categories are positive, demonstrating that PodAgent significantly outperforms directly prompting GPT-4 in generating podcast scripts across all evaluated dimensions.

\renewcommand{\arraystretch}{1.3}

\begin{table}[ht]
\caption{\centering \textbf{LLM-as-a-Judge:} comparison between GPT-4 and PodAgent (overall performance and 5 specific categories). Scores range from -3 to 3, where positive values favor PodAgent.}
\label{tab:llm-judge-part}
\centering
\setlength{\tabcolsep}{4pt} % Adjust column spacing
\resizebox{0.9\textwidth}{!}{
\begin{tabular}{lcccccc}
\toprule
\rowcolor[HTML]{F2F2F2} 
\textbf{Metrics} & \cellcolor{yellow!100}\textbf{Overall} & \textbf{Fiction} & \textbf{Education} & \textbf{Business} & \textbf{TrueCrime} & \textbf{Health \& Fitness} \\
\midrule
Coherence          & \cellcolor{yellow!10}0.7059 & 0.5000 & 0.8333 & 1.0000 & 1.0000 & 0.6667 \\
Engagingness       & \cellcolor{yellow!10}1.0294 & 1.1667 & 1.0000 & 1.1667 & 0.6667 & 1.1667 \\
Diversity          & \cellcolor{yellow!10}1.1765 & 1.3333 & 1.0000 & 1.3333 & 0.8333 & 1.5000 \\
Informativeness    & \cellcolor{yellow!10}1.6078 & 1.5000 & 1.6667 & 2.0000 & 1.1667 & 1.6667 \\
Speaker Difference & \cellcolor{yellow!10}1.0637 & 0.9167 & 1.0000 & 1.1667 & 0.6667 & 1.0000 \\
Overall            & \cellcolor{yellow!10}1.3064 & 1.2500 & 1.3333 & 1.6667 & 0.8333 & 1.2500 \\
\bottomrule
\end{tabular}}
\end{table}

%\vspace{-0.2cm}

\subsection{Speech-based evaluation}
\label{speech-eval}
%\vspace{-0.1cm}

To ensure fairness, all open-source TTS systems use the same PodAgent-generated scripts. Subjective tests use a spontaneous version from MoonCast, while objective evaluations use the original PodAgent scripts, as filler words in the spontaneous version challenge metrics like WER.

%\vspace{-0.3cm}

%\vspace{-0.3cm}

\begin{figure}[H]
  \centering      
  \includegraphics[width=1\columnwidth]{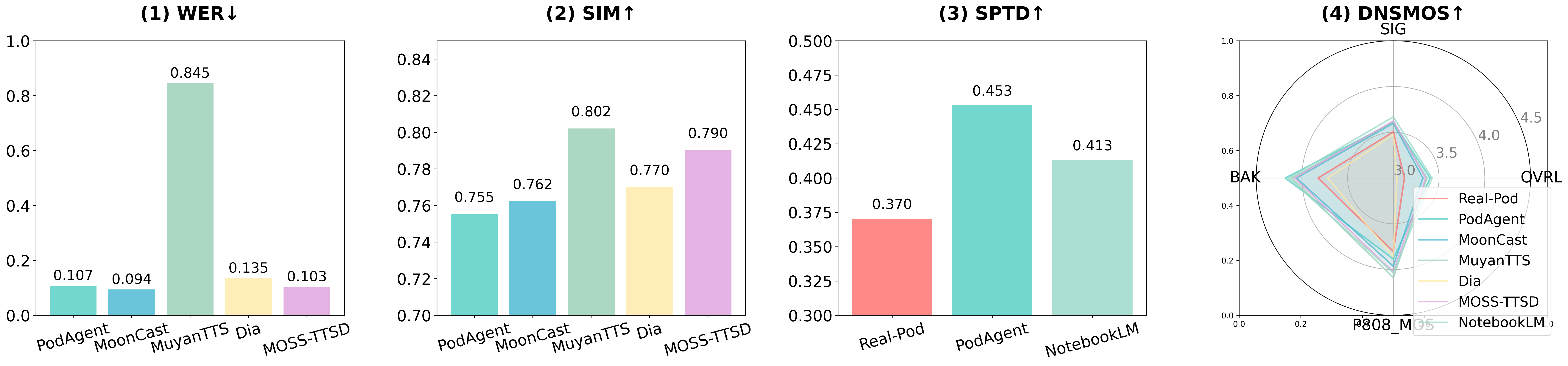}
  \captionsetup{justification=centering}
  \caption{\textbf{Speech-based evaluation:} objective metrics (WER, SPTD, SIM, DNSMOS).}
  \label{fig:speech_obj}
\end{figure}

\textbf{WER.} Figure \ref{fig:speech_obj}-(1) shows the WER results calculated for the entire conversation script. All systems, except MuyanTTS, achieve WER scores below 20\%. Analysis of sampled MuyanTTS outputs reveals robustness issues like repeated sentences and the insertion of unknown content.

\textbf{SIM.} The SIM metric evaluates zero-shot TTS systems' ability to replicate the timbre of a reference voice. PodAgent, MoonCast, MuyanTTS, Dia, and MOSS-TTSD—are assessed as shown in Figure \ref{fig:speech_obj}-(2). Each system uses the reference voice selected by PodAgent for the topic. The performance rankings are: MuyanTTS, MOSS-TTSD, Dia, MoonCast, and PodAgent. PodAgent's relatively low score in this metric likely stems from its instruction-following style control strategy. While this approach enhances overall conversational expressiveness, it can reduce speaker similarity.

\textbf{SPTD.} Figure \ref{fig:speech_obj}-(3) shows timbre variation across speakers in the conversation for three systems: Real-Pod, PodAgent, and NotebookLM. Real-Pod reflects real-world podcasts, PodAgent uses a voice selection mechanism for distinct voices, and NotebookLM fixed voices (one male, one female). The SPTD scores rank as follows: PodAgent, NotebookLM, and Real-Pod. This likely reflects that real-world podcasts prioritize guest expertise and availability over timbre differences. PodAgent demonstrates an effective automated voice selection process for podcast creation.

\textbf{DNSMOS.} The DNSMOS metric was applied to all systems to evaluate speech quality as in Figure \ref{fig:speech_obj}-(4). PodAgent, MoonCast, MuyanTTS, MOSS-TTSD, and NotebookLM achieve similar scores, while Real-Pod and Dia show noticeable declines in speech quality. For Real-Pod, the lower scores are due to: (1) real podcasts often use MSE for enhancement, requiring speech-MSE separation before evaluation, which may leave residual MSE artifacts, and (2) human-created podcasts involve recording, editing, or post-processing that introduce noise or instability. Dia struggles with long-form speech synthesis. Its outputs for lengthy podcast scripts frequently feature overly fast speaking speeds and occasional sentence truncations, leading to its relatively low DNSMOS performance.

\begin{table}[H]
\caption{Dialogue Naturalness Evaluation - statistical information for filtering.}
\label{tab:judger_metrics}
\vspace{-0.2cm}
\centering
\setlength{\tabcolsep}{5pt} % Adjust column spacing
\resizebox{0.9\textwidth}{!}{
\begin{tabular}{lcccccccccc}
\toprule
\rowcolor[HTML]{F2F2F2} 
\textbf{Judger} & \textbf{1} & \textbf{2} & \textbf{3} & \textbf{4} & \textbf{5} & \textbf{6} & \cellcolor{red!30}\textbf{7} & \textbf{8} & \textbf{9} & \textbf{10} \\
\midrule
LQ Last (\%)     & 94.12      & 100        & 100        & 100        & 100        & 100        & 100        & 100        & 100        & 100        \\
HQ Top2 (\%)     & 88.24      & 88.24      & 58.82      & 58.82      & 94.12      & 64.71      & \cellcolor{red!10}17.65      & 58.82      & 64.71      & 94.12      \\
\midrule
\rowcolor[HTML]{F2F2F2} 
\textbf{Judger} & \textbf{11} & \textbf{12} & \textbf{13} & \textbf{14} & \textbf{15} & \textbf{16} & \textbf{17} & \cellcolor{red!30}\textbf{18} & \textbf{19} & \cellcolor{red!30}\textbf{20} \\
\midrule
LQ Last (\%)     & 94.12       & 100         & 100         & 100         & 100         & 100         & 100         & \cellcolor{red!10}76.47       & 100         & 100         \\
HQ Top2 (\%)     & 94.12       & 82.35       & 64.71       & 82.35       & 88.24       & 58.82       & 64.71       & \cellcolor{red!10}47.06       & 52.94       & \cellcolor{red!10}35.29       \\
\bottomrule
\end{tabular}}
\end{table}

\vspace{-0.3cm}

\textbf{Dialogue Naturalness Evaluation.} We released the task on Prolific\footnote{\url{https://www.prolific.com/}}, requesting 20 native English-speaking participants from the US/UK. We set the filter rules as: 1) Over 90\% of LQ samples must be marked as the worst, as the synthesized samples from eSpeak are obviously robotic and unnatural. 2) Over 50\% of HQ samples must rank in the top-2 best. While it is possible for other systems to achieve a better score than the real podcast, the evaluation of the real podcast should also remain above average. Table \ref{tab:judger_metrics} presents the two statistical metrics for the submission results. Based on these rules, Judger-7, 18 and 20 can be excluded. We also provide the box plot for each Judger in Figure \ref{fig:speech-sbj-filter} in the appendix for more advanced analysis. For instance, apart from the LQ samples, Judger-20 assigns similar scores to all other systems, further confirming the invalidity of this submission.

%\vspace{-0.2cm}

\begin{figure}[H]
  \centering      
  \includegraphics[width=0.9\columnwidth]{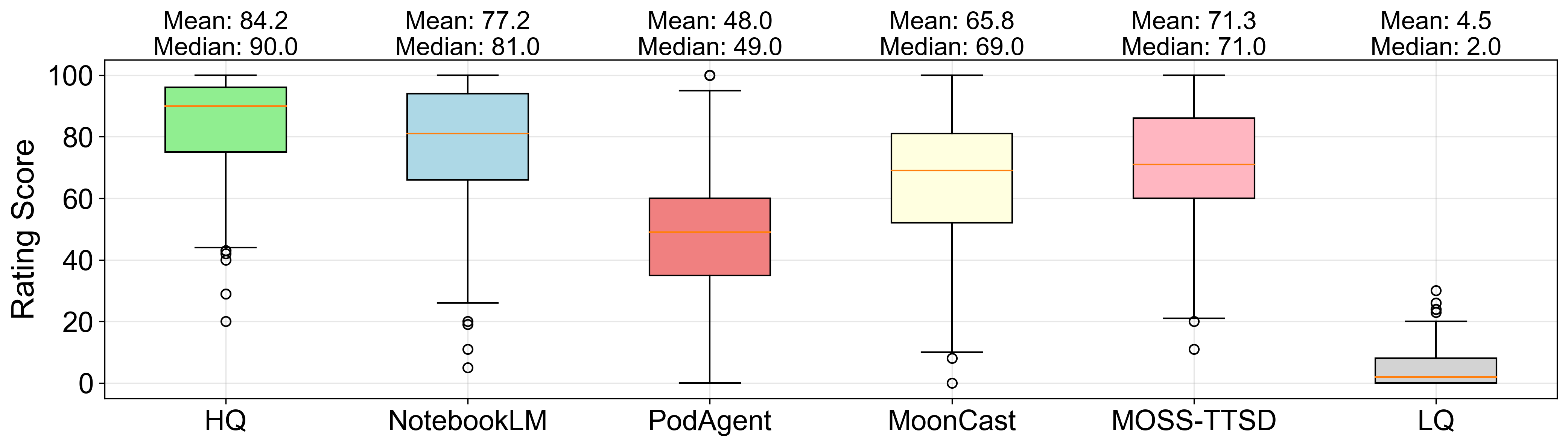}
  \captionsetup{justification=centering}
  \caption{Dialogue Naturalness Evaluation - overall result.}
  \label{fig:speech-sbj-merge}
\end{figure}

\textit{Result Analysis.} After excluding unqualified submissions, we analyzed system performance based on the remaining 17 valid submissions. Figure \ref{fig:speech-sbj-merge} presents the final results. We can observe that dialogue segments from real podcasts (HQ) achieved the highest scores, which aligns with expectations. NotebookLM, a closed-source product, ranked second, reflecting the high naturalness of its synthesized dialogue speech. Among the three open-source podcast generation systems, PodAgent scored the lowest, which is reasonable since its backend TTS system, CosyVoice2, is limited to single-sentence synthesis. In contrast, MoonCast and MOSS-TTSD, which support direct dialogue synthesis, performed better in dialogue naturalness evaluations. Overall, the evaluation results align with expectations, validating the rationality and effectiveness of our evaluation method design.

\vspace{-0.3cm}

%\vspace{-0.5cm}

\subsection{Audio-based evaluation}
\label{exp:audio}

\begin{figure}[H]
  \centering      
  \includegraphics[width=1\columnwidth]{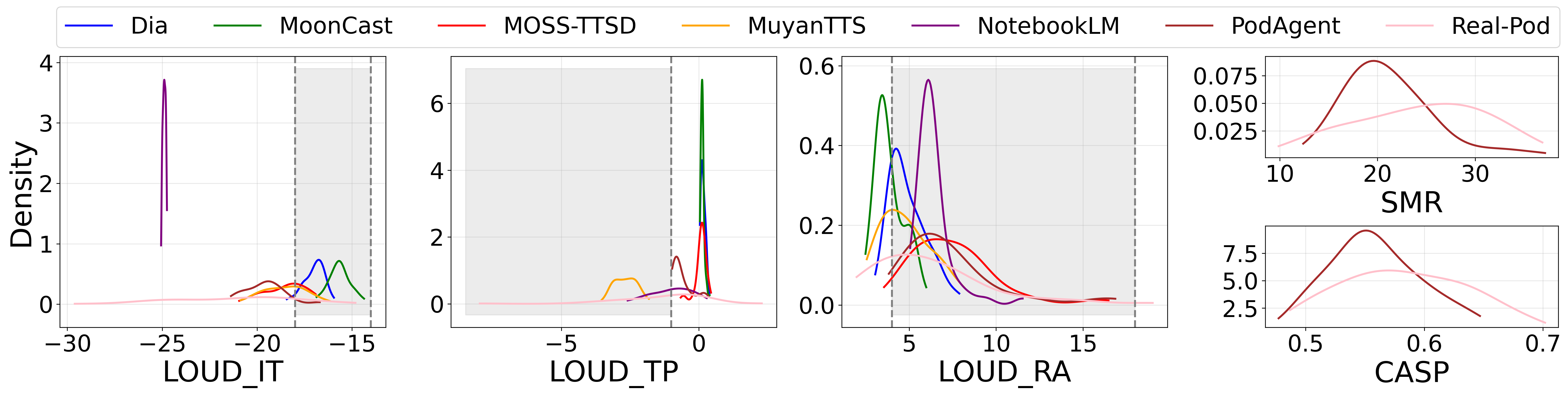}
  \captionsetup{justification=centering}
  \caption{Density distributions of audio-based objective metrics.}
  \label{fig:audio-obj}
\end{figure}

\textbf{Loudness.} Figure \ref{fig:audio-obj} presents the density distribution of loudness-related metrics, enabling a comparative analysis with the reference range. All seven systems are included. For \textbf{LOUD\_IT}, Dia and MoonCast align well with the reference range, while NotebookLM's loudness centers around -25. Real-Pod, as manually produced audio,  shows a highly scattered loudness distribution. For \textbf{LOUD\_TP}, Muyan-TTS performs best, with all samples maintaining a true peak loudness below -1. In contrast, MoonCast, Dia and MOSS-TTSD perform poorly, while Real-Pod continues to exhibit scattered results. For \textbf{LOUD\_RA}, MoonCast has a relatively narrow loudness variation range, while PodAgent and MOSS-TTSD display richer variance. Quantitative scores are detailed in Table \ref{tab:audio-obj-scores}.

\textbf{SMR and CASP.} PodAgent and Real-Pod are evaluated for these MSE-related metrics. From the density distribution in Figure \ref{fig:audio-obj}, PodAgent exhibits a more concentrated distribution compared to Real-Pod. For \textbf{SMR}, the SMR\_SCORE in Table \ref{tab:audio-obj-scores} shows that all PodAgent samples achieve an SMR greater than 0, whereas some Real-Pod cases fail to meet this requirement. For \textbf{CASP}, a higher score indicates better MSE-Speech harmony. Real-Pod demonstrates a higher upper limit, which is expected as exceptional human artistic creations naturally surpass AI-generated outputs. However, PodAgent delivers more consistent performance, and the overall gap between the two systems is not significant, making it an alternative way to enhance creative efficiency.

\begin{figure}[H]
  \centering      
  \includegraphics[width=1\columnwidth]{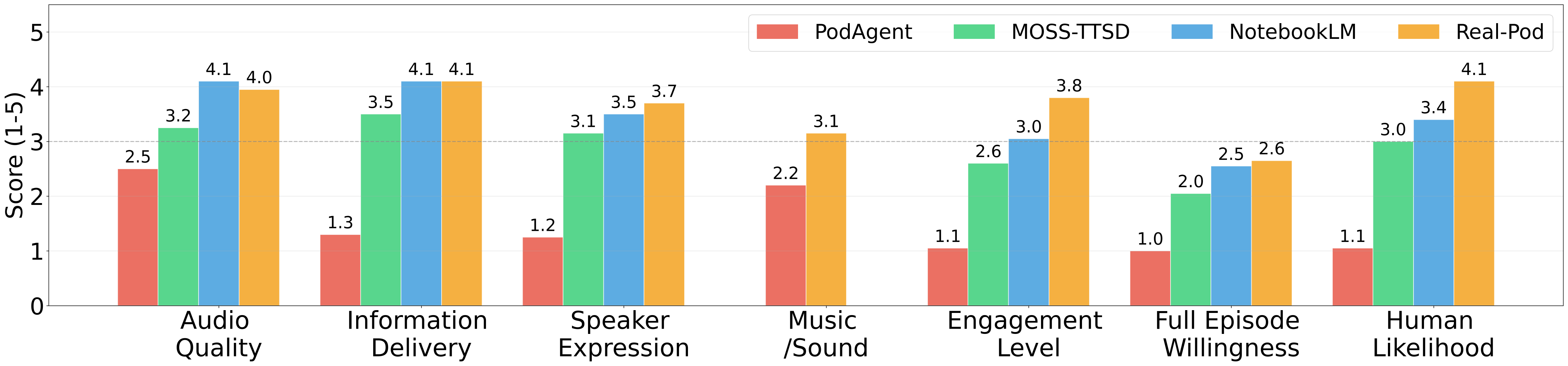}
  \captionsetup{justification=centering}
  \caption{Questionnaire-based MOS test}
  \label{fig:audio-sbj}
\end{figure}

\textbf{Questionnaire-based MOS Test.} We recruited native English speakers from Prolific for this test. Section \ref{audio} describes our final test design. Prior to this, we conducted a \textit{Pilot Test} using the questionnaire design in Figure \ref{fig:audio-sbj-pilot}. Based on feedback, we made the following improvements: 1) Reduced the scoring scale from 10 to 5 with clear definitions to reduce ambiguity and improve consistency. 2) Refined the questions to introduce perceptual and preference-based considerations. 3) Added a justification requirement for each question. These changes increased the pass rate from 75\% to 90\%. 

In addition to direct scores, we also derive a corresponding score based on users' justifications. Specifically, given justification texts from multiple systems for the same question, GPT-4 uses the following prompt to score: \textit{"For each system, summarize the corresponding comments into one sentence and assign a score between 1 and 5."} A detailed experiment setup is provided in Appendix \ref{appendix:Audio-sbj-final}, and separate scores are listed in Table \ref{tab:audio-sbj-QJ}. Figure \ref{fig:audio-sbj} shows the final results, averaging the direct score and the justification-based score. From the result, we can observe that:
%(Q.) represents the average score from the direct scoring answers, and (J.) represents the score derived from the justifications. 
\begin{itemize}[leftmargin=10pt]

\item \textit{Speech (naturalness and authenticity) is the most dominant factor affecting the listener's experience}.  In Section \ref{speech-eval}, PodAgent scored low in dialogue naturalness due to using a single-sentence synthesis TTS system, leading to consistently poor results in this MOS test. This outcome is expected, as dialogue speech is the core component of podcast-like audio programs. Although PodAgent's Music/Sound (harmony) score is below Real-Pod (consistent with the results of objective metric - CASP), it is significantly higher than its scores in other metrics, indicating that \textit{the gap between PodAgent and Real-Pod in music harmony is smaller than in speech naturalness}.
%than its other metrics
%the Dialogue Naturalness Evaluation in 

\item \textit{Real podcasts perform best in most metrics (5/7).} Real-Pod significantly outperforms other systems on holistic metrics like Engagement Level (EL) and Human Likelihood (HL). However, Full Episode Willingness (FEW) scores are low across all systems, with NotebookLM and Real-Pod scoring similarly. \textit{This highlights the value of perceptual and preference-based question design in the test.} FEW, a preference-based question, garnered justifications like ``the topic is not of interest to me" for lower scores. In contrast, higher scores for EL and HL indicate that users tend to exclude subjective factors (e.g., personal topic interest) when rating audio performance. A similar pattern is observed in Information Delivery (effectiveness) and Speaker Expression (preference).

\item In the Audio Quality metric, while PodAgent and MOSS-TTSD score lower than Real-Pod, PodAgent performs better here than in other metrics, and NotebookLM slightly surpasses Real-Pod. As noted, human-made podcasts often exhibit inconsistent audio quality due to complex production. User feedback, like ``Little bit of mic hiss/bloom but otherwise fine," supports this observation. This highlights that \textit{when conversational realism approaches that of real speech, AI-based methods offers an advantage in their controllability and consistency in producing high-quality audio}.
\end{itemize}

%\vspace{-0.2cm}

%\vspace{-0.5cm}

\section{Conclusion}
PodEval is the first comprehensive evaluation framework for podcast-like audio generation, tackling the challenges of assessing open-ended, long-form content. We constructed a real-world podcast dataset as a benchmark for human-level creative quality across diverse topics and formats. By decomposing evaluation into text, speech, and audio, PodEval introduced multidimensional methods combining objective metrics and well-designed subjective listening tests. Experiments with various podcast generation systems, including open-source, closed-source, and human-made examples, validated the framework's effectiveness. The results offer insights into the strengths and weaknesses of different systems (e.g. Figure \ref{fig:sys_report_podagent}), highlighting PodEval's role in advancing podcast generation research and inspiring future work on evaluating open-ended, long-form content generation task.
%underscoring PodEval's significance in advancing research on podcast generation and inspiring future work on evaluating open-ended, long-form content.
%advancing the research on long-form audio generation.

%By decomposing evaluation into three dimensions—text, speech, and audio—PodEval applies tailored methods combining objective metrics and carefully designed subjective listening tests to ensure data validity.

\section{Ethics statement}

This work introduces PodEval, a comprehensive framework for evaluating podcast-like audio generation, with careful consideration of ethical implications. The \textit{Real-Pod dataset} was constructed using publicly available podcasts in alignment with fair use, avoiding sensitive or private data. Instead of directly providing audio files, the dataset offers publicly accessible download links and download toolkit to reduce the risk of misuse and ensure proper attribution. \textit{Subjective evaluations} were conducted using crowdsourced workers recruited through the Prolific platform, with compensation exceeding the platform’s minimum wage requirements. Reliability was ensured through attention-check questions and clear instructions for participants. To mitigate bias, the framework incorporates \textit{diverse topics and evaluators}, promoting inclusivity and fairness. While PodEval aims to advance AI-assisted podcast generation, we emphasize its role as a tool to enhance, not replace, human creativity. PodEval is designed to foster innovation while adhering to principles of transparency, fairness, and ethical AI development.

\section{Reproducibility statement}

To ensure the reproducibility of our work, the PodEval framework is fully open-source and accessible at \url{https://github.com/yujxx/PodEval}. The repository contains all necessary datasets, scripts, and tools to replicate the experiments described in this paper.

\subsection*{How to Use the Repository}

\begin{enumerate}
    \item Clone the Repository.

    \item Set Up the Environment according to the README.
    
    \item Proccess dataset or Run Evaluations following the corresponding instructions.
\end{enumerate}

\subsection*{Repository Structure}

The repository is organized into the following directories:

\begin{itemize}[leftmargin=10pt]
    \item \textbf{\texttt{Real\_Pod/}}  
    \begin{itemize}
        \item Provides the \textit{Real-Pod dataset}, a curated collection of real-world podcast episodes. Includes 51 topics across 17 categories, representing diverse podcast scenarios.   
        \item See \texttt{Real\_Pod/README.md} for dataset preparation and usage instructions.
    \end{itemize}

    \item \textbf{\texttt{Text\_Eval/}}  
    \begin{itemize}
        \item Tools for \textit{text-based evaluation} of dialogue scripts. Includes both \textit{Quantitative Metrics} and \textit{LLM-as-a-Judge} methods.
        \item See \texttt{Text\_Eval/README.md} for instructions on running text evaluations.
    \end{itemize}

    \item \textbf{\texttt{Speech\_Audio\_Objective\_Evaluation/}}  
    \begin{itemize}
        \item Toolkit for \textit{objective evaluation} of podcast audio and speech quality. Includes DNSMOS, WER, SIM, SPTD, Loudness, SMR, and CASP.
        \item See \texttt{Speech\_Audio\_Obj\_Eval/README.md} for metric calculations and usage.
    \end{itemize}

    \item \textbf{\texttt{Subjective\_Listening\_Tests/}}  
    \begin{itemize}
        \item Framework for \textit{subjective human evaluations} of podcast speech and audio. One is \textit{Dialogue Naturalness Evaluation} and the other one is \textit{Questionnaire-based MOS Test}.
        \item See \texttt{Subjective\_Listening\_Tests/README.md} for test setup and implementation details. We also provide \textit{website demo} \href{https://podeval.github.io/PodEval-Subjective/?config=dialogue.yaml}{link1} \href{https://podeval.github.io/PodEval-Subjective/?config=questionnaire.yaml}{link2} that allow users to intuitively view the test design and participate it.
    \end{itemize}
\end{itemize}

By following the provided instructions and leveraging the structured tools within each directory, users can reproduce all experiments and adapt \textbf{PodEval} for further research.

\bibliography{iclr2026_conference}
\bibliographystyle{iclr2026_conference}

\appendix
\clearpage

\raggedbottom
\section{Appendix}

\subsection{Use of large language models}
Large Language Models (LLMs) utilized in this work are as follows:(1) \textit{Topics Initiation} during data processing of the Real-Pod dataset, which is elaborated in Section~\ref{dataset}. (2) \textit{LLM-as-a-Judge} method in text-based evaluation, which is illustrated in Section~\ref{text}. (3) \textit{Summarized Users' Justifications} in the Questionnaire-based MOS Test, which is described in Section~\ref{exp:audio} (Questionnaire-based MOS Test).

\subsection{Text-based evaluation}

\subsubsection{Quantitative metrics}
\label{appendix:quant_text}

\begin{table}[h]
\caption{\textbf{GPT-4}: Quantitative metrics in text-based evaluation.}
\label{tab:gpt4_quant_text_sorted}
\centering
\setlength{\tabcolsep}{5pt} % 调整列间距
\begin{tabular}{lcccccc}
\toprule
\rowcolor[HTML]{F2F2F2} 
\textbf{Metrics} & \cellcolor{yellow!100}\textbf{Overall} & \textbf{Fiction} & \textbf{Education} & \textbf{Business} & \textbf{True Crime} & \textbf{Health \& Fitness} \\
\midrule
Distinct\_2      & \cellcolor{yellow!10}0.9619 & 0.9643 & 0.9588 & 0.9567 & 0.9689 & 0.9638 \\
Info-Dens        & \cellcolor{yellow!10}6.4507 & 6.5865 & 6.4569 & 6.3213 & 6.6541 & 6.3880 \\
Sem-Div          & \cellcolor{yellow!10}0.1293 & 0.1204 & 0.1115 & 0.1214 & 0.1443 & 0.1106 \\
MATTR            & \cellcolor{yellow!10}0.6914 & 0.7027 & 0.6989 & 0.6933 & 0.6831 & 0.6870 \\
\midrule
\rowcolor[HTML]{F2F2F2} 
\textbf{Metrics} & \textbf{Sports} & \textbf{Comedy} & \textbf{History} & \textbf{News} & \textbf{TV \& Film} & \textbf{Society \& Culture} \\
\midrule
Distinct\_2      & 0.9536 & 0.9633 & 0.9471 & 0.9486 & 0.9678 & 0.9659 \\
Info-Dens        & 6.4228 & 6.2256 & 6.3792 & 6.3225 & 6.7614 & 6.6473 \\
Sem-Div          & 0.1248 & 0.1356 & 0.1451 & 0.1208 & 0.1553 & 0.1507 \\
MATTR            & 0.6973 & 0.6922 & 0.6905 & 0.6756 & 0.6903 & 0.6901 \\
\midrule
\rowcolor[HTML]{F2F2F2} 
\textbf{Metrics} & \textbf{Arts} & \textbf{Leisure} & \textbf{Music} & \textbf{Kids} & \textbf{Mental Health} & \textbf{Science \& Tech} \\
\midrule
Distinct\_2      & 0.9675 & 0.9729 & 0.9555 & 0.9559 & 0.9699 & 0.9710 \\
Info-Dens        & 6.5054 & 6.5233 & 6.4119 & 6.2310 & 6.4787 & 6.3454 \\
Sem-Div          & 0.1374 & 0.1117 & 0.1320 & 0.1229 & 0.1247 & 0.1286 \\
MATTR            & 0.6885 & 0.7136 & 0.6677 & 0.6884 & 0.6994 & 0.6960 \\
\bottomrule
\end{tabular}
\end{table}

\begin{table}[H]
\caption{\textbf{PodAgent}: Quantitative metrics in text-based evaluation.}
\label{tab:podagent_quant_text_sorted}
\centering
\setlength{\tabcolsep}{5pt} % 调整列间距
\begin{tabular}{lcccccc}
\toprule
\rowcolor[HTML]{F2F2F2} 
\textbf{Metrics} & \cellcolor{yellow!100}\textbf{Overall} & \textbf{Fiction} & \textbf{Education} & \textbf{Business} & \textbf{True Crime} & \textbf{Health \& Fitness} \\
\midrule
Distinct\_2      & \cellcolor{yellow!10}0.9741 & 0.9743 & 0.9730 & 0.9758 & 0.9796 & 0.9825 \\
Info-Dens        & \cellcolor{yellow!10}7.1767 & 7.3791 & 7.2163 & 7.1126 & 7.1810 & 7.2927 \\
Sem-Div          & \cellcolor{yellow!10}0.1372 & 0.1384 & 0.1210 & 0.1254 & 0.1514 & 0.1171 \\
MATTR            & \cellcolor{yellow!10}0.7216 & 0.7399 & 0.7291 & 0.7258 & 0.7263 & 0.7386 \\
\midrule
\rowcolor[HTML]{F2F2F2} 
\textbf{Metrics} & \textbf{Sports} & \textbf{Comedy} & \textbf{History} & \textbf{News} & \textbf{TV \& Film} & \textbf{Society \& Culture} \\
\midrule
Distinct\_2      & 0.9678 & 0.9808 & 0.9483 & 0.9735 & 0.9782 & 0.9744 \\
Info-Dens        & 7.1239 & 7.1600 & 7.1004 & 7.1282 & 7.3311 & 6.9568 \\
Sem-Div          & 0.1487 & 0.1236 & 0.1543 & 0.1379 & 0.1690 & 0.1344 \\
MATTR            & 0.7183 & 0.7248 & 0.6752 & 0.7156 & 0.7274 & 0.7119 \\
\midrule
\rowcolor[HTML]{F2F2F2} 
\textbf{Metrics} & \textbf{Arts} & \textbf{Leisure} & \textbf{Music} & \textbf{Kids} & \textbf{Mental Health} & \textbf{Science \& Tech} \\
\midrule
Distinct\_2      & 0.9701 & 0.9790 & 0.9739 & 0.9747 & 0.9815 & 0.9725 \\
Info-Dens        & 7.1977 & 7.3227 & 7.0558 & 7.0930 & 7.1822 & 7.1711 \\
Sem-Div          & 0.1283 & 0.1445 & 0.1440 & 0.1353 & 0.1259 & 0.1331 \\
MATTR            & 0.7101 & 0.7275 & 0.7114 & 0.7279 & 0.7328 & 0.7249 \\
\bottomrule
\end{tabular}
\end{table}

\begin{table}[H]
\caption{\textbf{MoonCast}: Quantitative metrics in text-based evaluation.}
\label{tab:mooncast_quant_text_sorted}
\centering
\setlength{\tabcolsep}{5pt} % 调整列间距
\begin{tabular}{lcccccc}
\toprule
\rowcolor[HTML]{F2F2F2}
\textbf{Metrics} & \cellcolor{yellow!100}\textbf{Overall} & \textbf{Fiction} & \textbf{Education} & \textbf{Business} & \textbf{True Crime} & \textbf{Health \& Fitness} \\
\midrule
Distinct\_2 & \cellcolor{yellow!10}0.9128 & 0.9219 & 0.8998 & 0.8952 & 0.9132 & 0.9478 \\
Info-Dens & \cellcolor{yellow!10}6.0935 & 6.3613 & 5.9779 & 5.9931 & 6.0388 & 6.4230 \\
Sem-Div & \cellcolor{yellow!10}0.1326 & 0.1515 & 0.1079 & 0.1326 & 0.1405 & 0.1324 \\
MATTR & \cellcolor{yellow!10}0.6323 & 0.6598 & 0.6237 & 0.6310 & 0.6391 & 0.6698 \\
\midrule
\rowcolor[HTML]{F2F2F2}
\textbf{Metrics} & \textbf{Sports} & \textbf{Comedy} & \textbf{History} & \textbf{News} & \textbf{TV \& Film} & \textbf{Society \& Culture} \\
\midrule
Distinct\_2 & 0.9159 & 0.9232 & 0.9169 & 0.9047 & 0.9408 & 0.8959 \\
Info-Dens & 6.1933 & 6.1729 & 6.2229 & 5.9672 & 6.2855 & 5.8031 \\
Sem-Div & 0.1451 & 0.1311 & 0.1460 & 0.1176 & 0.1318 & 0.1282 \\
MATTR & 0.6402 & 0.6435 & 0.6276 & 0.6111 & 0.6595 & 0.6121 \\
\midrule
\rowcolor[HTML]{F2F2F2}
\textbf{Metrics} & \textbf{Arts} & \textbf{Leisure} & \textbf{Music} & \textbf{Kids} & \textbf{Mental Health} & \textbf{Science \& Tech} \\
\midrule
Distinct\_2 & 0.9252 & 0.8889 & 0.8957 & 0.9039 & 0.9222 & 0.9073 \\
Info-Dens & 6.2335 & 5.9713 & 5.9411 & 5.9291 & 6.0298 & 6.0459 \\
Sem-Div & 0.1444 & 0.1309 & 0.1227 & 0.1291 & 0.1187 & 0.1432 \\
MATTR & 0.6370 & 0.6124 & 0.6035 & 0.6183 & 0.6321 & 0.6277 \\
\bottomrule
\end{tabular}
\end{table}

\begin{table}[H]
\caption{\textbf{Real-Pod}: Quantitative metrics in text-based evaluation.}
\label{tab:real_quant_text_sorted}
\centering
\setlength{\tabcolsep}{5pt} % 调整列间距
\begin{tabular}{lcccccc}
\toprule
\rowcolor[HTML]{F2F2F2} 
\textbf{Metrics} & \cellcolor{yellow!100}\textbf{Overall} & \textbf{Fiction} & \textbf{Education} & \textbf{Business} & \textbf{True Crime} & \textbf{Health \& Fitness} \\
\midrule
Distinct\_2      & \cellcolor{yellow!10}0.9200 & 0.9292 & 0.9275 & 0.9049 & 0.9169 & 0.9273 \\
Info-Dens        & \cellcolor{yellow!10}8.1168 & 8.2849 & 8.1160 & 7.7755 & 8.5675 & 7.9301 \\
Sem-Div          & \cellcolor{yellow!10}0.1677 & 0.1776 & 0.1579 & 0.1433 & 0.1906 & 0.1646 \\
MATTR            & \cellcolor{yellow!10}0.6251 & 0.6313 & 0.6346 & 0.6041 & 0.6261 & 0.6380 \\
\midrule
\rowcolor[HTML]{F2F2F2} 
\textbf{Metrics} & \textbf{Sports} & \textbf{Comedy} & \textbf{History} & \textbf{News} & \textbf{TV \& Film} & \textbf{Society \& Culture} \\
\midrule
Distinct\_2      & 0.9244 & 0.8994 & 0.9272 & 0.9100 & 0.9201 & 0.8932 \\
Info-Dens        & 8.0993 & 8.2755 & 8.8282 & 7.7886 & 8.4005 & 7.7375 \\
Sem-Div          & 0.1919 & 0.1660 & 0.1845 & 0.1618 & 0.1784 & 0.1701 \\
MATTR            & 0.6434 & 0.5999 & 0.6304 & 0.6102 & 0.6363 & 0.5823 \\
\midrule
\rowcolor[HTML]{F2F2F2} 
\textbf{Metrics} & \textbf{Arts} & \textbf{Leisure} & \textbf{Music} & \textbf{Kids} & \textbf{Mental Health} & \textbf{Science \& Tech} \\
\midrule
Distinct\_2      & 0.9111 & 0.9242 & 0.9420 & 0.9092 & 0.9298 & 0.9439 \\
Info-Dens        & 8.1093 & 7.6949 & 8.0925 & 7.7708 & 8.2119 & 8.3031 \\
Sem-Div          & 0.1653 & 0.1591 & 0.1761 & 0.1492 & 0.1668 & 0.1485 \\
MATTR            & 0.6063 & 0.6176 & 0.6513 & 0.6200 & 0.6373 & 0.6582 \\
\bottomrule
\end{tabular}
\end{table}

\subsubsection{LLM-as-a-Judge}
\label{appendix:LLM_text}

\begin{table}[H]
\caption{ \textbf{LLM-as-a-Judge: comparison between GPT-4 and PodAgent.} Scores range from -3 to 3. Positive values indicate that PodAgent outperforms GPT-4; Negative values suggest the opposite.}
\label{tab:appendix_text_llm}
\centering
\setlength{\tabcolsep}{5pt} % 调整列间距
\resizebox{\textwidth}{!}{
\begin{tabular}{lcccccc}
\toprule
\rowcolor[HTML]{F2F2F2} 
\textbf{Metrics} & \cellcolor{yellow!100}\textbf{Overall} & \textbf{Fiction} & \textbf{Education} & \textbf{Business} & \textbf{True Crime} & \textbf{Health \& Fitness} \\
\midrule
Coherence          & \cellcolor{yellow!10}0.7059 & 0.5000 & 0.8333 & 1.0000 & 1.0000 & 0.6667 \\
Engagingness       & \cellcolor{yellow!10}1.0294 & 1.1667 & 1.0000 & 1.1667 & 0.6667 & 1.1667 \\
Diversity          & \cellcolor{yellow!10}1.1765 & 1.3333 & 1.0000 & 1.3333 & 0.8333 & 1.5000 \\
Informativeness    & \cellcolor{yellow!10}1.6078 & 1.5000 & 1.6667 & 2.0000 & 1.1667 & 1.6667 \\
Speaker Difference & \cellcolor{yellow!10}1.0637 & 0.9167 & 1.0000 & 1.1667 & 0.6667 & 1.0000 \\
Overall            & \cellcolor{yellow!10}1.3064 & 1.2500 & 1.3333 & 1.6667 & 0.8333 & 1.2500 \\
\midrule
\rowcolor[HTML]{F2F2F2} 
\textbf{Metrics} & \textbf{Sports} & \textbf{Comedy} & \textbf{History} & \textbf{News} & \textbf{TV \& Film} & \textbf{Society \& Culture} \\
\midrule
Coherence          & 0.5000 & 0.8333 & 1.1667 & 0.6667 & 0.8333 & 0.1667 \\
Engagingness       & 1.1667 & 1.5000 & 1.5000 & 0.6667 & 0.1667 & 0.6667 \\
Diversity          & 1.1667 & 1.8333 & 1.5000 & 1.3333 & 1.3333 & 0.8333 \\
Informativeness    & 1.5000 & 2.1667 & 1.5000 & 2.0000 & 1.3333 & 0.8333 \\
Speaker Difference & 1.1667 & 1.5000 & 1.1667 & 1.3333 & 1.1667 & 1.3333 \\
Overall            & 1.5000 & 1.8333 & 1.5000 & 1.5000 & 0.8333 & 0.5000 \\
\midrule
\rowcolor[HTML]{F2F2F2} 
\textbf{Metrics} & \textbf{Arts} & \textbf{Leisure} & \textbf{Music} & \textbf{Kids} & \textbf{Mental Health} & \textbf{Science \& Tech} \\
\midrule
Coherence          & 0.6667 & 0.5000 & 0.6667 & 0.5000 & 0.3333 & 1.1667 \\
Engagingness       & 1.1667 & 1.1667 & 1.1667 & 1.0000 & 0.8333 & 1.3333 \\
Diversity          & 1.1667 & 1.1667 & 1.0000 & 1.0000 & 0.3333 & 1.3333 \\
Informativeness    & 1.8333 & 2.0000 & 1.8333 & 1.3333 & 1.1667 & 1.8333 \\
Speaker Difference & 1.3333 & 1.1667 & 0.8333 & 0.8333 & 0.6667 & 0.8333 \\
Overall            & 1.5000 & 1.6667 & 1.5000 & 1.1667 & 0.8333 & 1.5417 \\
\bottomrule
\end{tabular}}
\end{table}

\clearpage
\subsection{Speech-based evaluation (Subjective)}
\label{apdx:speech-sbj}

\begin{figure}[h]
  \centering      
  \includegraphics[width=0.97\columnwidth]{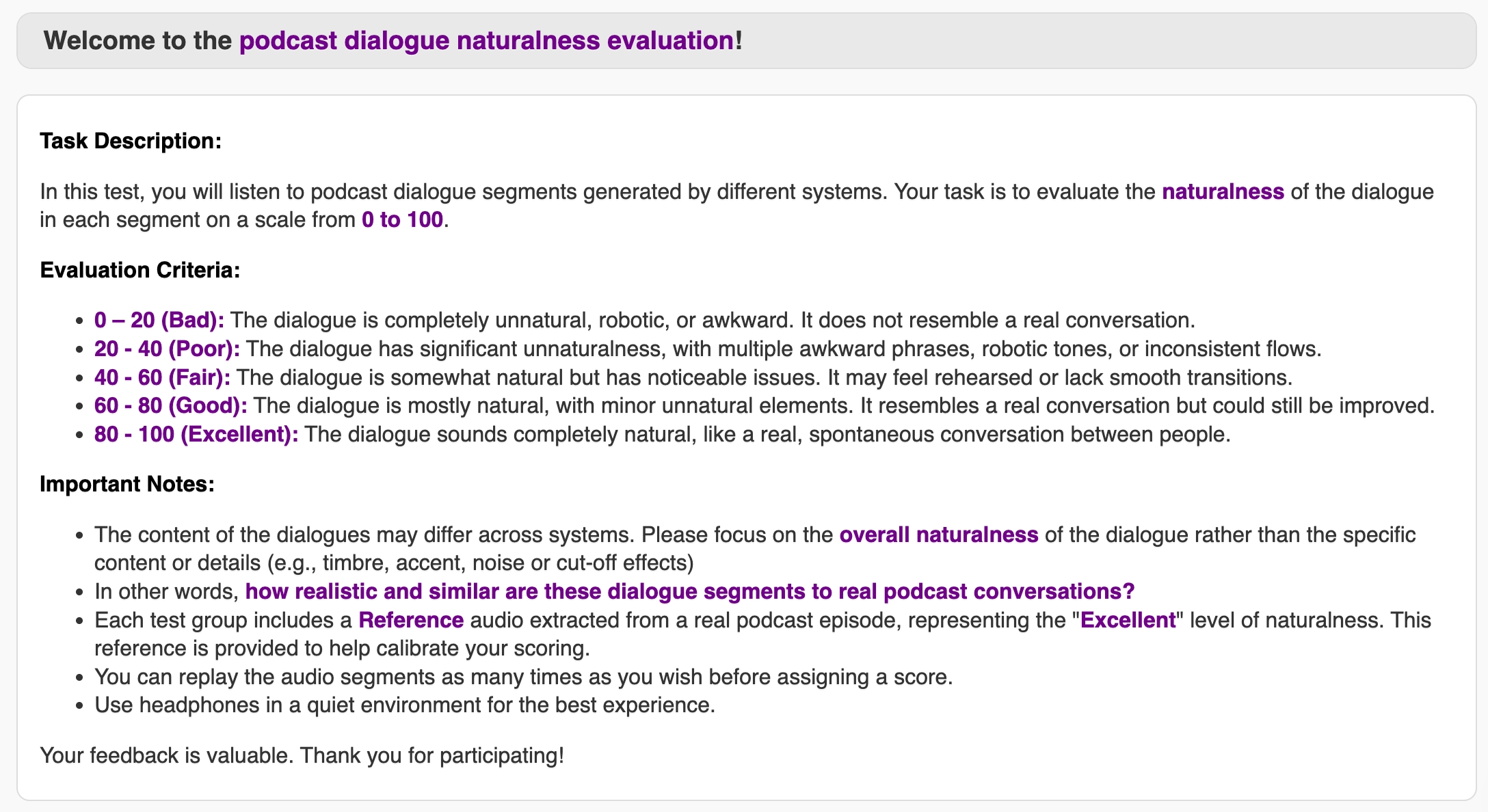}
  \captionsetup{justification=centering}
  \caption{Dialogue Naturalness Evaluation - Instruction page. }
  \label{fig:dia-inst}
\end{figure}

\begin{figure}[h]
  \centering      
  \includegraphics[width=0.97\columnwidth]{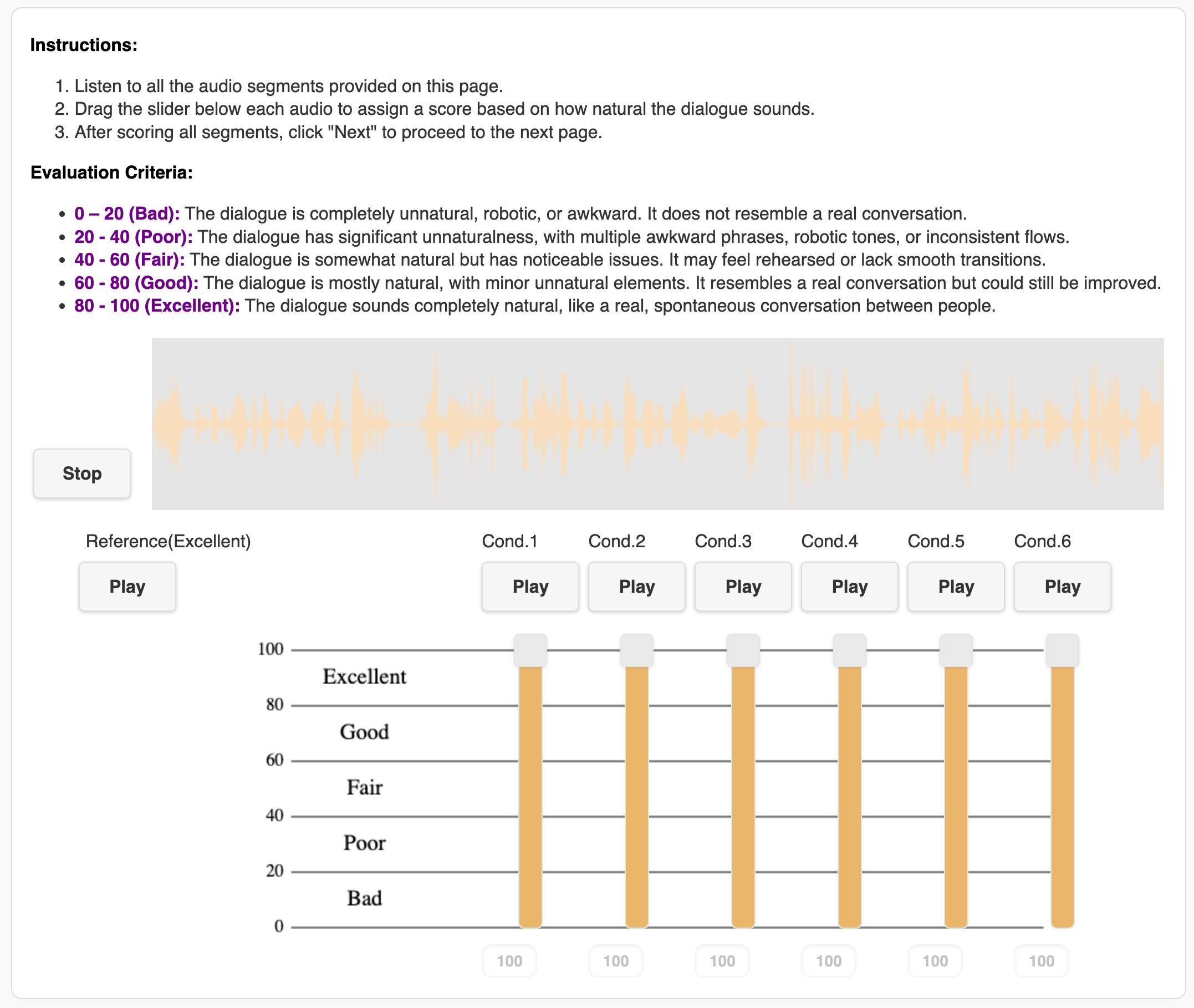}
  \captionsetup{justification=centering}
  \caption{Dialogue Naturalness Evaluation - Test page. }
  \label{fig:dia-test}
\end{figure}

\begin{figure}[h]
  \centering      
  \includegraphics[width=1\columnwidth]{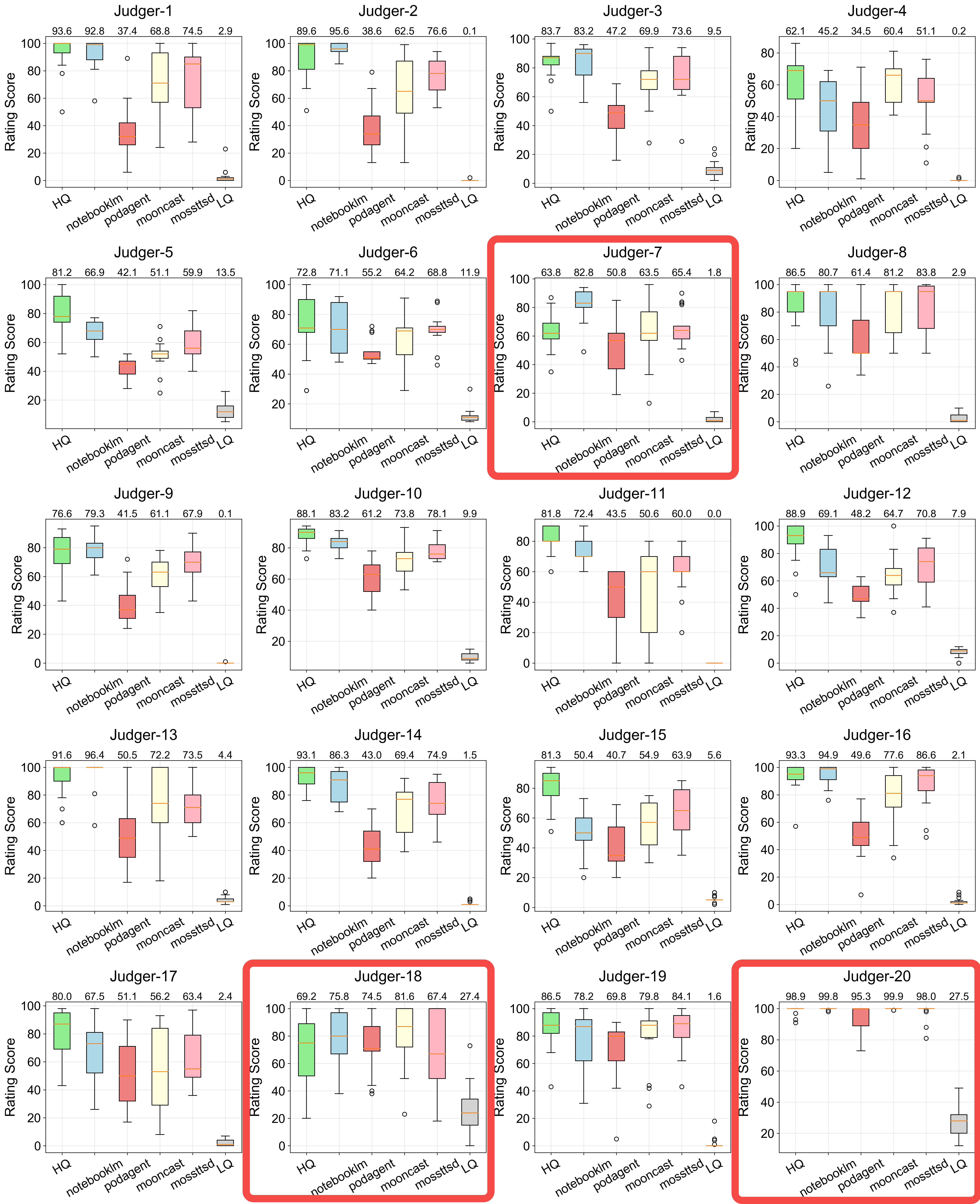}
  \captionsetup{justification=centering}
  \caption{Dialogue Naturalness Evaluation test results from each juders.}
  \label{fig:speech-sbj-filter}
\end{figure}

\clearpage
\subsection{Audio-based evaluation (objective)}
\label{ap_obj}

IDL: LOUD-IT; TP: LOUD-TP; LRA: LOUD-RA.

\begin{equation}
S_{\text{IDL}} = 
\begin{cases} 
1, & -18 \leq \text{IDL} \leq -14, \\ 
e^{-k_1 \cdot (-18 - \text{IDL})}, & \text{IDL} < -18, \\
e^{-k_2 \cdot (\text{IDL} + 14)}, & -14 < \text{IDL}, 
\end{cases}
\end{equation}

where \( k_1 \) is set as \( 0.0858 \) to ensure \( S_{\text{IDL}} \) is around \( 0.6 \) when \( \text{IDL} = -23 \), and \( k_2 \) is set as \( 0.3291 \) to make \( S_{\text{IDL}} \) close to \( 0 \) when \( \text{IDL} = 0 \).

\begin{equation}
S_{\text{TP}} = 
\begin{cases} 
1, & \text{TP} \leq -1 \\ 
e^{-k_3 \cdot (\text{TP} + 1)}, & \text{TP} > -1
\end{cases}
\end{equation}

where \( k_3 \) is set as \( 4.605 \) to ensure \( S_{\text{TP}} \) is close to \( 0 \) when \( \text{TP} \) approaches \( 0 \).

\begin{equation}
S_{\text{LRA}} = 
\begin{cases} 
1, & 4 \leq \text{LRA} \leq 18, \\ 
e^{-k_4 \cdot (4 - \text{LRA})}, & \text{LRA} < 4, \\ 
e^{-k_5 \cdot (\text{LRA} - 18)}, & \text{LRA} > 18.
\end{cases}
\end{equation}

where \( k_4 \) is set as \( 1.1513 \) to ensure \( S_{\text{LRA}} \) approaches \( 0 \) when \( \text{LRA} = 0 \), and \( k_5 \) is set as \( 0.2554 \) to ensure \( S_{\text{LRA}} \approx 0.6 \) when \( \text{LRA} = 20 \).

\begin{table}[ht]
\caption{Audio-based objective metrics - Quantitative scores.}
\label{tab:audio-obj-scores}
\centering
\setlength{\tabcolsep}{4pt} % Adjust column spacing
\resizebox{\textwidth}{!}{
\begin{tabular}{lccccc}
\toprule
\rowcolor[HTML]{F2F2F2} 
\textbf{System} & \textbf{LOUD\_IT\_SCORE} & \textbf{LOUD\_TP\_SCORE} & \textbf{LOUD\_LRA\_SCORE} & \textbf{SMR\_BASIC\_SCORE} & \textbf{CASP} \\
\midrule
Real-Pod   & 0.72  & 0.53  & 0.82  & 0.99  & 0.58 \\
PodAgent   & 0.80  & 0.32  & 1.00  & 1.00  & 0.56 \\
MoonCast   & 1.00  & 0.01  & 0.68  & -     & -    \\
Muyan-TTS  & 0.88  & 1.00  & 0.83  & -     & -    \\
Dia        & 0.98  & 0.01  & 0.95  & -     & -    \\
MOSS-TTSD  & 0.88  & 0.02  & 0.99  & -     & -    \\
NotebookLM & 0.51  & 0.56  & 1.00  & -     & -    \\
\bottomrule
\end{tabular}}
\end{table}

\clearpage
\subsection{Audio-based evaluation (subjective)}
\label{appendix:Audio-sbj}

\subsubsection{Pilot test}
\label{appendix:Audio-sbj-pilot}

\begin{figure}[h]
  \centering      
  \includegraphics[width=0.75\columnwidth]{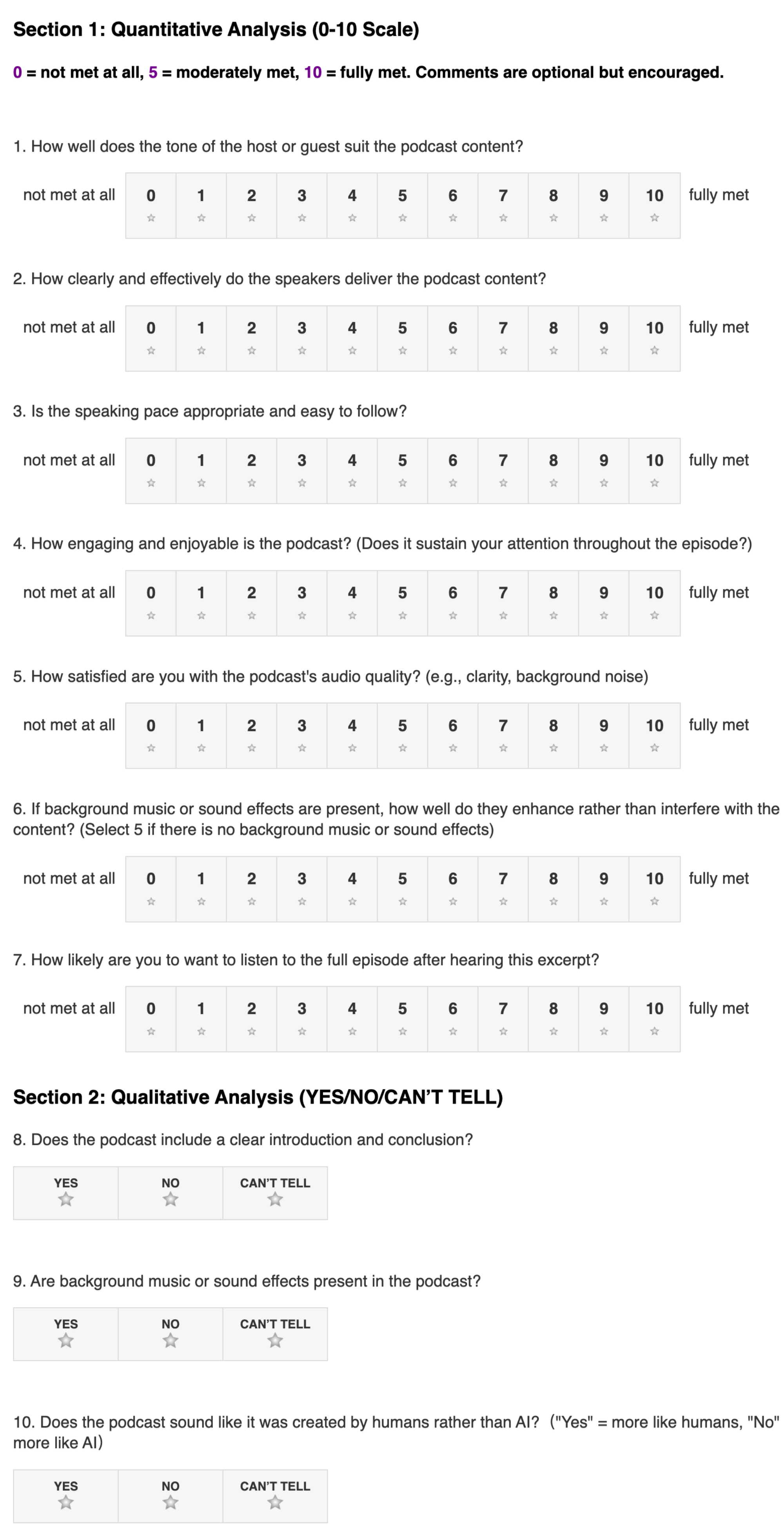}
  \captionsetup{justification=centering}
  \caption{Questionnaire-based MOS test - Pilot test version.}
  \label{fig:audio-sbj-pilot}
\end{figure}

\subsubsection{Questionnaire-based MOS test}
\label{appendix:Audio-sbj-final}

\textbf{Experiment Settings:} Lengthy listening tests can be exhausting and may lead to inaccurate feedback. It is essential to ensure the overall test duration does not exceed 30 minutes. In the Questionnaire-based MOS Test, each audio sample is around 3 minutes and requires answering 10 questions with corresponding justifications. Based on the Dialogue Naturalness Test results shown in Figure \ref{speech-eval}, we selected 4 representative systems. Each test group included four podcast samples from different systems but within the same podcast category. According to actual test results, each group took an average of 24 minutes to complete. The 4 representative systems are:

\begin{itemize}[leftmargin=10pt]
\item \textbf{PodAgent:} An open-source podcast generation framework incorporating conversation script generation, automatic voice selection, speech synthesis, and BMSE enhancement.
\item \textbf{MOSS-TTSD:} Achieved the highest score among the open-source systems utilized in the Dialogue Naturalness Evaluation (Figure \ref{speech-eval}).
\item \textbf{NotebookLM:} A pioneering podcast generation product, widely recognized for its exceptional performance, is nearly indistinguishable from real podcasts.
\item \textbf{Real-Pod:} A collection of podcasts sourced from the real world.
\end{itemize}

\begin{figure}[h]
  \centering      
  \includegraphics[width=1\columnwidth]{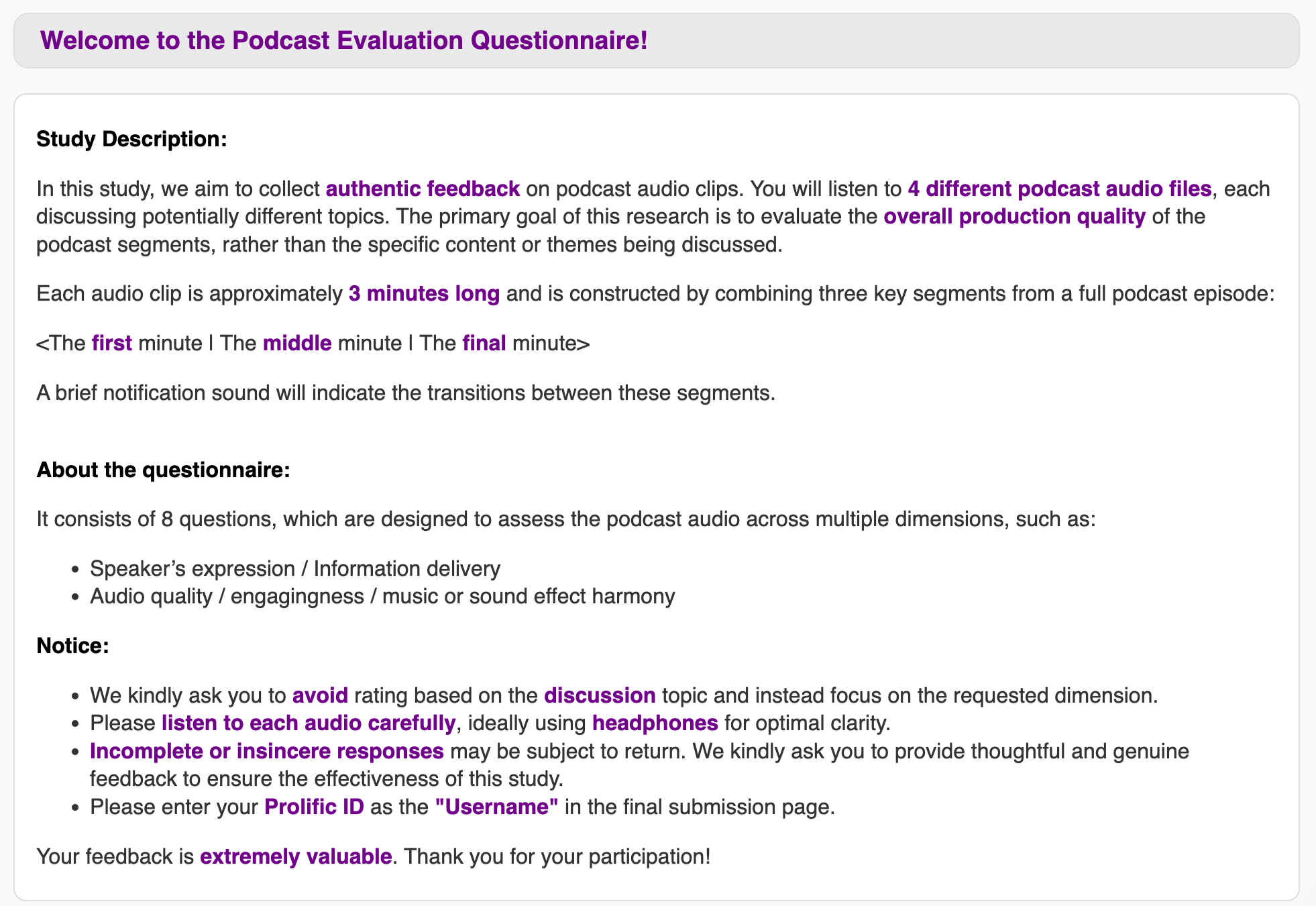}
  \captionsetup{justification=centering}
  \caption{Questionnaire-based MOS test - Final version - Instruction page.}
  \label{fig:audio-sbj-final-inst}
\end{figure}

\clearpage

\begin{figure}[h]
  \centering      
  \includegraphics[width=1\columnwidth]{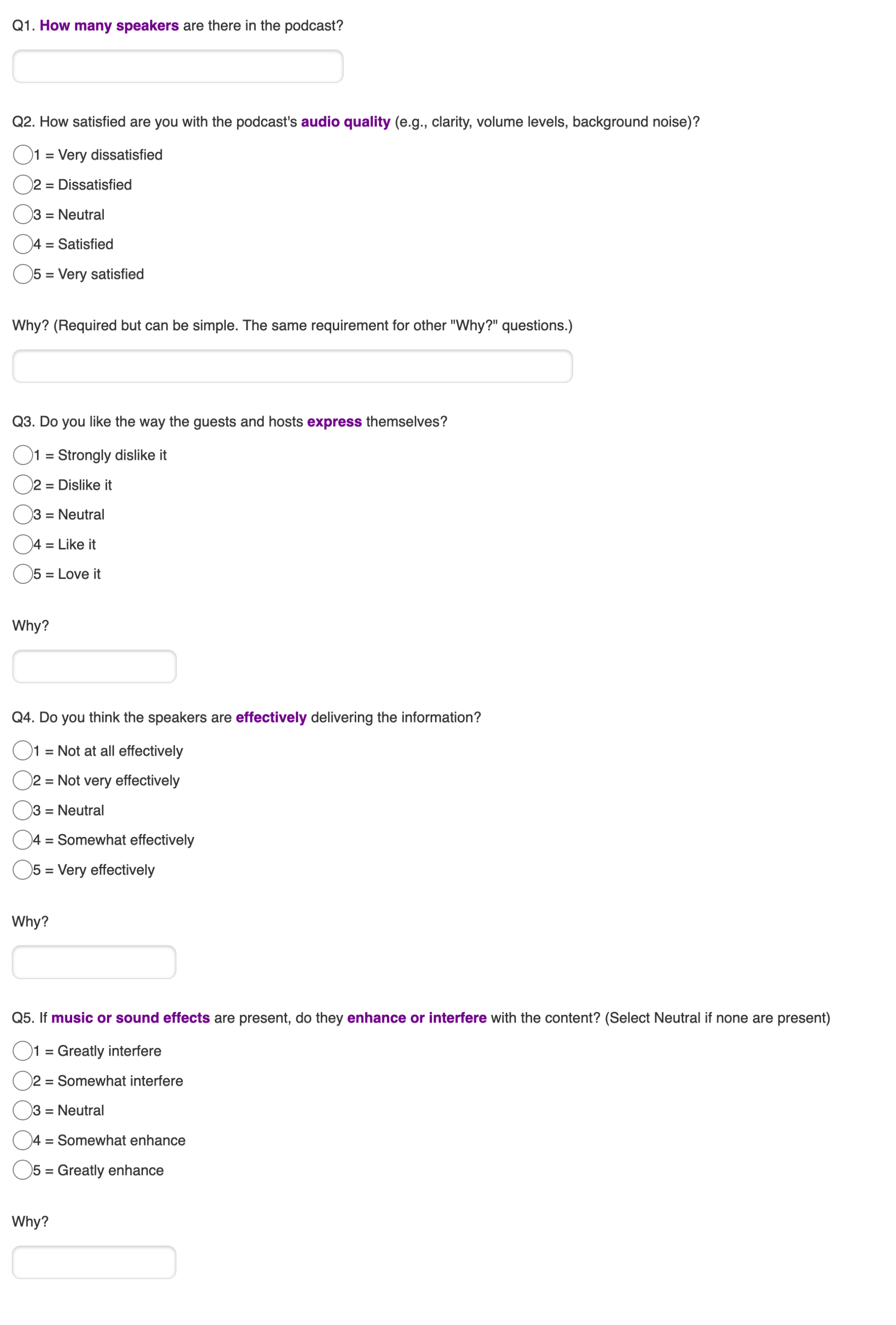}
  \captionsetup{justification=centering}
  \caption{Questionnaire-based MOS test - Final version (Question 1-5).}
  \label{fig:audio-sbj-final-1}
\end{figure}

\begin{figure}[h]
  \centering      
  \includegraphics[width=1\columnwidth]{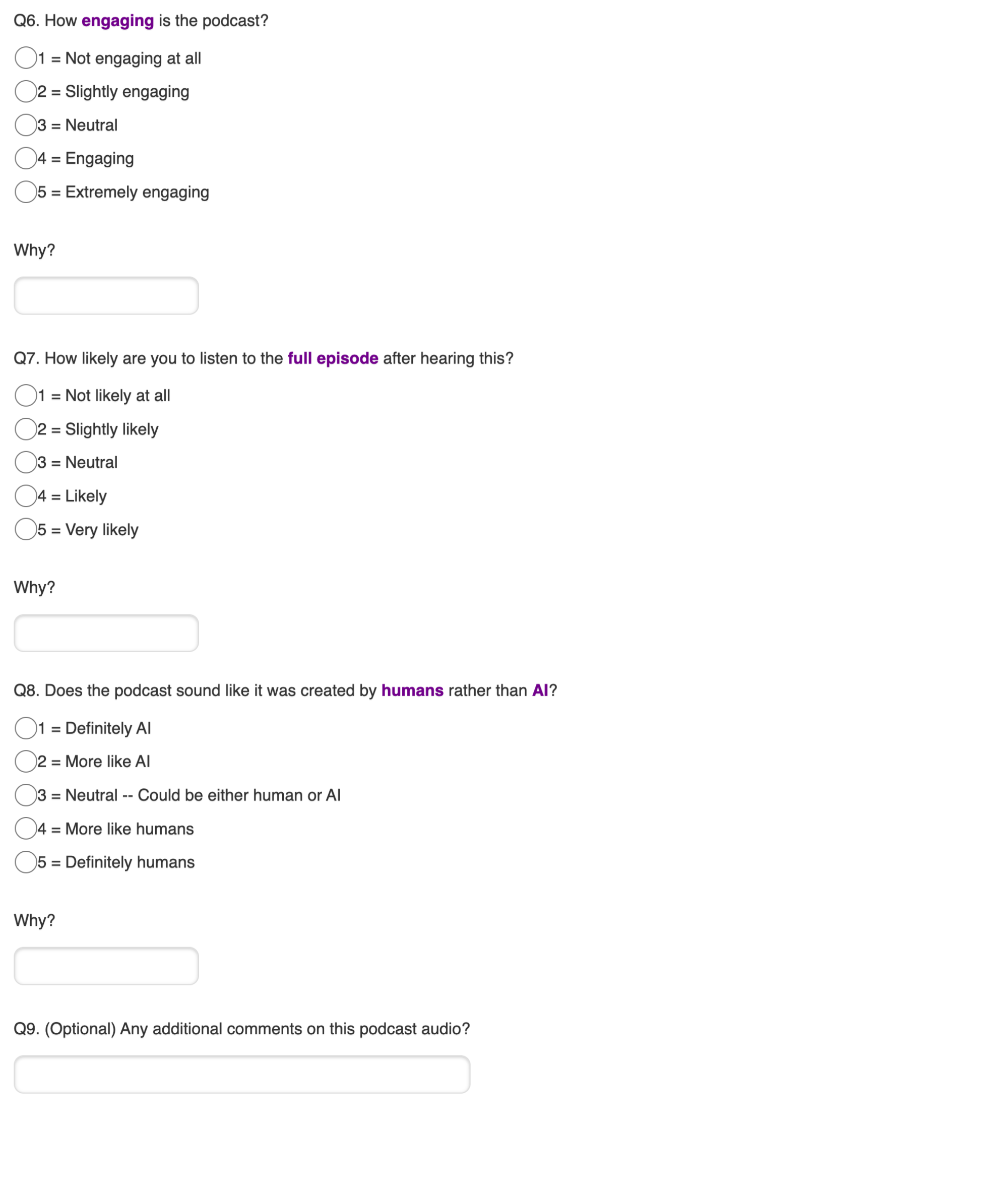}
  \captionsetup{justification=centering}
  \caption{Questionnaire-based MOS test - Final version (Question 6-9).}
  \label{fig:audio-sbj-final-2}
\end{figure}

\clearpage

\begin{table}[h]
\caption{Questionnaire-based MOS test - (Q.) represents the average score from the direct scoring answers, and (J.) represents the score derived from the justifications. }
\label{tab:audio-sbj-QJ}
\centering
%\resizebox{0.8\textwidth}{!}{
\begin{tabular}{lcccccccc}
\toprule
\diagbox{Metrics}{Systems} & 
\multicolumn{2}{c}{\textbf{MOSS-TTSD}} & 
\multicolumn{2}{c}{\textbf{NotebookLM}} & 
\multicolumn{2}{c}{\textbf{PodAgent}} & 
\multicolumn{2}{c}{\textbf{Real-Pod}} \\
\cmidrule(lr){2-3} \cmidrule(lr){4-5} \cmidrule(lr){6-7} \cmidrule(lr){8-9}
 & \textbf{Q.} & \textbf{J.} & \textbf{Q.} & \textbf{J.} & \textbf{Q.} & \textbf{J.} & \textbf{Q.} & \textbf{J.} \\
\midrule
Information Delivery     & 4.0 & 3.0 & 4.2 & 4.0 & 1.6 & 1.0 & 4.2 & 4.0 \\
Music/Sound Effects      & N/A & N/A & N/A & N/A & 2.4 & 2.0 & 3.3 & 3.0 \\
Engagement Level         & 2.2 & 3.0 & 3.1 & 3.0 & 1.1 & 1.0 & 3.6 & 4.0 \\
Full Episode Likelihood  & 2.1 & 2.0 & 2.1 & 3.0 & 1.0 & 1.0 & 2.3 & 3.0 \\
Human Likelihood         & 3.0 & 3.0 & 3.3 & 3.5 & 1.1 & 1.0 & 4.2 & 4.0 \\
Audio Quality            & 3.5 & 3.0 & 4.2 & 4.0 & 3.0 & 2.0 & 3.9 & 4.0 \\
Speaker Expression       & 3.3 & 3.0 & 4.0 & 3.0 & 1.5 & 1.0 & 3.4 & 4.0 \\
\bottomrule
\end{tabular}
\end{table}

\subsection{System analysis report}
\label{apd:sys_report}

\begin{figure}[h]
  \centering      
  \includegraphics[width=0.9\columnwidth]{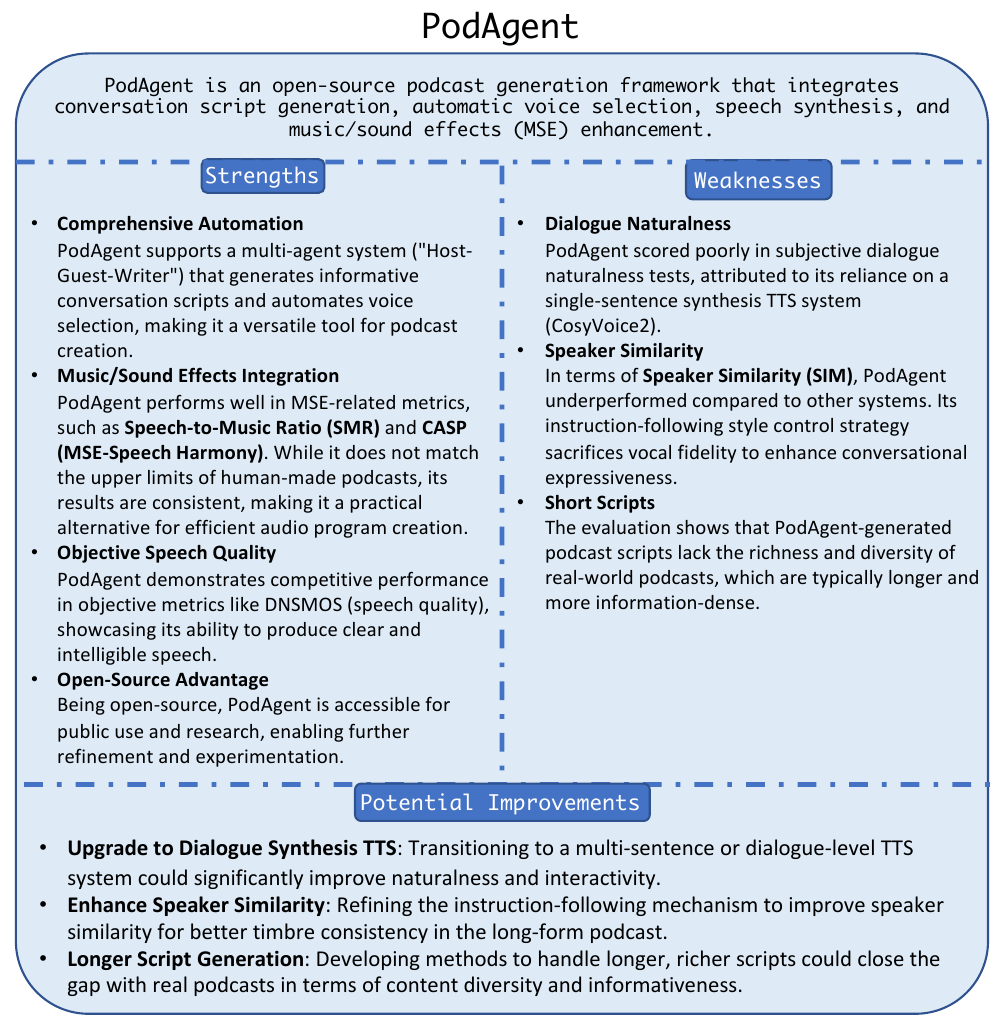}
  \captionsetup{justification=centering}
  \caption{System analysis report based on PodEval - PodAgent.}
  \label{fig:sys_report_podagent}
\end{figure}

\end{document}